\documentclass{appolb}
\usepackage{epsfig}
\usepackage{amsmath}
\usepackage{graphicx}
\usepackage{psfrag}
\usepackage{amsfonts}
\usepackage{amssymb}

\def\runhead{~}

\begin{document}

\title{\hfill TPJU-1/2003\\~~ \\Pion Generalized Distribution
Amplitudes in the Nonlocal Chiral Quark Model}
\author{\textbf{Micha{\l}
Prasza\l{}owicz{\thanks{e-mail: michal@if.uj.edu.pl}}
 and
Andrzej Rostworowski{\thanks{e-mail: arostwor@th.if.uj.edu.pl}}
 }\\
\emph{M.Smoluchowski Institute of Physics,} \\
\emph{Jagellonian University,} \\
\emph{Reymonta 4, 30-059 Krak\'{o}w, Poland.}}
\date{\today}
\maketitle
\begin{abstract}
We use a simple, instanton motivated, nonlocal chiral quark model
to calculate pion Generalized Distribution Amplitudes (GDAs). The
nonlocality appears due to the momentum dependence of the
constituent quark mass, which we take in a form of a generalized
dipole formula.  With this choice all calculations can be performed
directly in the Minkowski space and the sensitivity to the shape
of the cutoff function can be studied. We demonstrate that the
model fulfills soft pion theorems both for chirally even and
chirally odd GDAs. The latter one cannot be derived by the methods
of current algebra. Whenever possible we compare our results with
the existing data. This can be done for the pion electromagnetic
form factor and quark distributions as measured in the proton-pion
scattering.

\end{abstract}


\newpage

\section{Introduction}

%
%
\label{intro}
%
%

In this work we calculate two pion distribution amplitudes
($2\pi$DAs) and pion skewed parton distributions ($\pi$SPDs) in
the instanton motivated, effective chiral quark model. $2\pi$DAs
and $\pi$SPDs are defined as Fourier transforms of matrix elements
of certain light-cone operators, taken between the pion states.
$2\pi$DAs appear in the amplitude for the process
$\gamma^{\ast}\gamma\,\rightarrow\,\pi\pi$, if the virtuality of
the photon is much larger then the squared invariant mass of the
two pion system \cite{DGPT}. $2\pi$DAs describe the transition
from partons to two pions in a final state. They are also related
to the pion electromagnetic form factor in the time like region
and pion electromagnetic radius.

The process $\gamma^{\ast}\gamma\,\rightarrow\,\pi\pi$ can be
related by crossing symmetry to virtual Compton scattering
$\gamma^{\ast}\pi \,\rightarrow\,\gamma\pi$. This process
factorizes into a hard photon-parton scattering and $\pi$SPDs
\cite{Ji}--\nocite{R}\cite{RadAPPB}. In the forward limit the
latter reduce to the quark densities which are measurable in the
$\pi-$p Drell-Yan process or prompt photon production
\cite{SMRS,GRVS}. Finally certain integral of $\pi$SPDs gives pion
electromagnetic form factor in the space like region.

Pions are the simplest hadronic states being $q\bar{q}$ pairs and
Goldstone bosons of spontaneously broken chiral symmetry at the
same time. Therefore their properties may be calculated with
little dynamical input, relying on their chiral structure and
chiral symmetry breaking. In this work we use the instanton
motivated effective chiral quark model \cite{PP} with nonlocal
interactions. This model has been successfully applied to the
calculation of the leading twist pion distribution amplitude
\cite{PP}--\nocite{Bochum,PR}\cite{PRcond}, $2\pi$DAs
\cite{PW}--\nocite{P}\cite{PW2} and pion SPDs \cite{P,PW2}. The main ingredient of the
model is the momentum dependence of the constituent quark mass
which regularizes certain, otherwise divergent, integrals. To make
the calculation feasible this momentum dependence has been taken
in the form \cite{PR} :
\begin{equation}
M_{k}=M\left(
\frac{-\Lambda^{2}}{k^{2}-\Lambda^{2}+i\epsilon}\right)
^{2n}=M F^{2}(k). \label{M}%
\end{equation}
$M$ is the constituent quark mass at zero momentum. Its value,
obtained from the \textit{gap} equation, is approximately $350$
MeV. Quantities $n$ and $\Lambda=\Lambda(n;M)$ are  model
parameters. As explained below we expect the model to be roughly
independent of $n$, if the value of $\Lambda$ is properly
adjusted. The formula (\ref{M}) maybe thought to be instanton
motivated in a sense, that when continued to Euclidean momenta
($k^{2}\,\rightarrow\,-k_{E}^{2}$), it reproduces reasonably well
\cite{PR} the momentum dependence calculated explicitly
\cite{CMcD, DP} in the instanton model of the QCD vacuum:
\begin{equation}
F_{\mbox{{\small inst}}}(k_{E})=
2z[I_{0}(z)K_{1}(z)-I_{1}(z)K_{0}(z)]-2I_{1}(z)K_{1}(z),
\label{Minst}%
\end{equation}
where $z=\rho k_{E}/2$, with $\rho=(600~\mbox{MeV})^{-1}$ being an
average instanton size.

The calculations presented here may be viewed as an extension of
the previous results of Refs.\cite{PW}--\nocite{P}\cite{PW2} and
were partially reported in \cite{Moriond,BadH}. Our motivations
are both theoretical and phenomenological. First, the asymptotics
of Eq.(\ref{Minst}) suggests $n=3/2$ in Eq.(\ref{M}), however,
taking the non-integer $n$ results in additional difficulties. In
the previous works \cite{PP,Bochum,PW}--\nocite{P}\cite{PW2} the
value $n=1$ was assumed. Therefore it should be checked that the
results do not depend strongly on $n$. This is indeed true for the
pion distribution amplitude, as has been shown in \cite{PR}.
Secondly in \cite{PP, Bochum,PW}--\nocite{P}\cite{PW2} the momentum
dependence of the constituent mass in the quark propagators was
neglected. With the method developed in \cite{PR} we can calculate
generalized parton distributions for arbitrary $n$, keeping track
of momentum dependence of the constituent quark mass both in
numerators and denominators even for the convergent quantities.
This is in fact necessary if one insists on the soft pion theorems
which would be otherwise not fulfilled. Apart from a well known
soft pion theorem for a chirally even $2\pi$DA \cite{P}, we derive
(in the framework of the present model) a soft pion theorem for a
chirally odd $2\pi$DA.

On the phenomenological side, apart from the calculation of the
pion generalized distribution amplitudes (GDAs for short),
\emph{i.e.} $2\pi$DAs and $\pi$SPDs themselves, we calculate pion
electromagnetic radius, electromagnetic form factor and quark
distributions, and compare them with the existing data. This
comparison although not bad, is not entirely satisfactory. We
shall discuss our results in Sect.\ref{sect:sum}. We start with
the short description of the model in Sect.\ref{model} Then, in
Sect.\ref{def  &  prop}, we proceed with definitions of $2\pi$DAs
and $\pi $SPDs and their general properties. In
Sect.\ref{model_GDA} we sketch the calculations and present
numerical results. In Sect.\ref{softpions} we demonstrate soft
pion theorems. Finally, in Sect.\ref{strucfun}, we discuss pion
quark densities calculated in the model and compare them with data
and other theoretical calculations. Technical details can be found
in the Appendix.

\section{Effective chiral quark model}
\label{model}

For two quark flavors ($u$ and $d$) the effective model we use to calculate
pion GDAs is given by the action (in momentum space):
\begin{equation}
S_{eff}=\int\frac{d^{4}k}{(2\pi)^{4}}\,\bar{\psi}(k)\rlap{/}k\,\psi
(k)-\int\frac{d^{4}p}{(2\pi)^{4}}\frac{d^{4}k}{(2\pi)^{4}}\,\bar{\psi}%
(p)\sqrt{M_{p}}U^{\gamma_{5}}(p-k)\sqrt{M_{k}}\psi(k), \label{S}%
\end{equation}
where
\[
\psi(x)=\int\frac{d^{4}k}{(2\pi)^{4}}e^{-i\,k x}\psi(k).
\]
The matrix
\begin{equation}
U^{\gamma_{5}}(x)=\exp\left[  \frac{i}{F_{\pi}}\gamma^{5}\tau^{a}\pi
^{a}(x)\right]  =1+\frac{i}{F_{\pi}}\gamma^{5}\tau^{a}\pi^{a}(x)-\frac
{1}{2F_{\pi}^{2}}\pi^{a}(x)\pi^{a}(x)+\ldots\label{U}%
\end{equation}
(in coordinate space) gives the interaction between quarks and
pions. There is no kinetic term for the pions and the pion field
is treated as an external source. $F_{\pi}=93$ MeV is the pion
decay constant, $\tau^{a}$ are Pauli matrices and the pion
$\hat{T}_{3}$ eigen states are: $\pi^{0}=\pi^{3}$,
$\pi^{+}=-\left(  \pi^{1}+i\pi^{2}\right)  /\sqrt{2}$,
$\pi^{-}=\left( \pi^{1}-i\pi^{2}\right)  /\sqrt{2}$. The momentum
dependence of the quark constituent mass is given by Eq.(\ref{M}).
The $\Lambda$ parameter is fixed once for all by adjusting the
value of $F_{\pi}(\Lambda)$ calculated in the effective model
(\ref{S}) to its physical value $F_{\pi}=93$ MeV. This has been
done in \cite{PR}. For example for $M=350$ MeV and $n=1$ we have
obtained  $\Lambda=1157$ MeV.

Neither $M$ nor $\Lambda$ should be identified with the
normalization scale $Q_{0}$ of the quantities calculated in the
model. The precise definition of $Q_{0}$ is only possible within
QCD and in all effective models one can use only qualitative
\emph{order of magnitude} arguments to estimate $Q_{0}$. Arguments
can be given that the characteristic scale of the instanton model
is of the order of the inverse instanton size $1/\rho = 600$~MeV,
or so. We shall come back to this question in Sect.\ref{strucfun}.

\section{Definitions and general properties of GDAs}
\label{def  &  prop}
%
%

We define two light-like vectors: $n^{\mu}=(1,0,0,-1)$ and $\tilde{n}^{\mu
}=(1,0,0,1)$. These vectors define plus and minus components of any
four-vector: $k^{+}\equiv k^{0}+k^{3}=n\cdot k$ and $k^{-}\equiv k^{0}%
-k^{3}=\tilde{n}\cdot k$.

\subsection{Two pion distribution amplitudes}
\label{sect:2pi}

We take the definitions of pion generalized distribution
amplitudes from \cite{P,PW2}.
\begin{align}
&  \int\frac{d\lambda}{2\pi}\exp\left(  -iu\lambda n\cdot P\right)
\left\langle \pi^{a}(p_{1})\pi^{b}(p_{2})\left|  \bar{\psi}^{f^{\prime}%
}(\lambda n)\Gamma\psi^{f}(0)\right|  0\right\rangle
\nonumber \\
&=\delta^{ab}\,\delta^{ff^{\prime}}\,\Phi_{2\pi}^{I=0}(u,v,s)+i\varepsilon
^{abc}\,(\tau^{c})^{ff^{\prime}}\,\Phi_{2\pi}^{I=1}(u,v,s),
\label{2piDA_def}
\end{align}
where $P=p_{1}+p_{2}$ is the total momentum of the two pion system. For the
leading twist (chirally even) $2\pi$DAs we have:
\begin{equation}
\Gamma=\rlap{/}n,
\label{Gamma_even}
\end{equation}
whereas for the chirally odd $2\pi$DAs
\begin{equation}
\Gamma=in^{\alpha}P^{\beta}\sigma_{\alpha\beta}.
\label{Gamma_odd}
\end{equation}
Note that with the definition (\ref{Gamma_odd}), chirally odd
$2\pi$DAs have the dimension of mass.
$2\pi$DAs depend on the following kinematical variables: squared
invariant mass of the two pions, $s=P^{2}$; the longitudinal
momentum fraction carried by quark with respect to the total
longitudinal momentum, $u=k^{+}/P^{+}$ and the longitudinal
momentum fraction carried by one of the pions with respect to the
total longitudinal momentum, $v=p_{1}^{+}/P^{+}$.

$2\pi$DAs have the following symmetries \cite{P}:
\begin{equation}
\Phi_{2\pi}^{I=0}(u,v,s)=-\Phi_{2\pi}^{I=0}(1-u,v,s)=\Phi_{2\pi}%
^{I=0}(u,1-v,s), \label{I02piDA_sym}%
\end{equation}%
\begin{equation}
\Phi_{2\pi}^{I=1}(u,v,s)=\Phi_{2\pi}^{I=1}(1-u,v,s)=-\Phi_{2\pi}%
^{I=1}(u,1-v,s). \label{I12piDA_sym}%
\end{equation}
The isovector chirally even $2\pi$DA is normalized to the pion
electromagnetic form factor in the time-like region:
\begin{equation}
\int\limits_{0}^{1}du\,\Phi_{2\pi}^{I=1}(u,v,s)=(1-2v)F_{\pi}^{\mbox{{\small
em}}}(s). \label{F_pi(s)}%
\end{equation}
From its $s$ dependence the pion electromagnetic radius can be evaluated
\begin{equation}
\langle(r_{\pi}^{em})^{2}\rangle=6\left.  \frac{dF_{\pi}^{\mbox{{\small
em}}}(s)}{ds}\right|  _{s=0}. \label{r_pi_2piDA}%
\end{equation}
The normalization condition for the isoscalar $2\pi$DA gives the
fraction of a pion momentum carried by the quarks,
$M_{2}^{(\pi)}$:
\begin{equation}
\int\limits_{0}^{1}du\,(2u-1)\Phi_{2\pi}^{I=0}(u,v,s=0)=-2v(1-v)M_{2}^{(\pi)}.
\label{2piDAI0_norm}%
\end{equation}

It is convenient do expand the $2\pi$DAs in the basis of the eigen
functions
of the ERBL equation \cite{LB, ER} (Gegenbauer polynomials $C_{n}^{3/2}%
(2u-1)$) and partial waves of scattered pions (Legendre
polynomials $P_{l}(1-2v)$):
\begin{equation}
\Phi_{2\pi}^{I}(u,v,s)=6u(1-u)\sum_{n=0}^{\infty}\sum_{l=0}^{n+1}B_{nl}%
^{I}(s)C_{n}^{3/2}(2u-1)P_{l}(1-2v). \label{2piDA_exp}%
\end{equation}
Because of the symmetry properties (\ref{I02piDA_sym},
\ref{I12piDA_sym}), the sum in  Eq.(\ref{2piDA_exp}) goes over odd
(even) $n$ and even (odd) $l$ for isoscalar (isovector) $2\pi$DA.

\begin{figure}[h]
\begin{center}
\psfrag{s}{$s=0$~GeV$^2$} \psfrag{u}{~} \psfrag{v}{~}
\psfrag{quark}{\footnotesize quark $u$} %
\psfrag{pion}{\footnotesize pion $v$}
\includegraphics[scale=0.6]{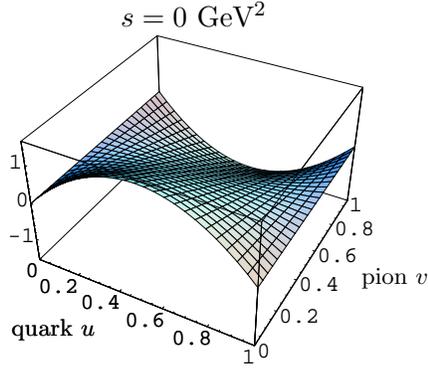}
\end{center}
\caption{{\footnotesize Asymptotic form the isovector 2$\pi$DA as
given
by Eq.(\ref{2piasymp}). }}%
\label{asymp}%
\end{figure}

The expansion coefficients $B_{nl}^I$ are renormalized
multiplicatively:
\begin{equation}
B_{nl}^I (s; Q) = B_{nl}^I (s; Q_0) \left( \frac {\alpha_s (Q)}
{\alpha_s (Q_0)} \right)^{\gamma_n}, \label{renormalizacja_B}
\end{equation}
with $\gamma_n$ being anomalous dimensions \cite{SV}. From
(\ref{2piDA_exp}) and (\ref{renormalizacja_B}) it is easy to read
the asymptotic form of $2\pi$DAs:
\begin{equation}
\Phi_{2 \pi}^{\mbox{\small as.} \, I=0} (u, v, s) = 0
\end{equation}
and
\begin{equation}
\Phi_{2 \pi}^{\mbox{\small as.} \, I=1} (u, v, s) = 6 u (1-u) (1 -
2v) B_{01}^{I=1} (s) \label{2piasymp}.
\end{equation}
This asymtotic form is plotted in Fig.\ref{asymp}. For chirally
even $2\pi$DA, from (\ref{F_pi(s)}), we have $B_{01}^{I=1} (s) =
F_{\pi}^{\mbox{\small em}} (s)$.

%
%
%
%

In Ref.\cite{P} the following soft pion theorem for the chirally
even $2\pi$DAs has been demonstrated. In the chiral limit
($m_{\pi}=0$), in the case of one of the pions momenta going to
zero ($v(1-v)\rightarrow0$) we get
\begin{equation}
\Phi_{2\pi}^{I=0}(u,v=0,s=0)=\Phi_{2\pi}^{I=0}(u,v=1,s=0)=0 \label{I0soft_pi}%
\end{equation}
and
\begin{equation}
\Phi_{2\pi}^{I=1}(u,v=0,s=0)=-\Phi_{2\pi}^{I=1}(u,v=1,s=0)=\phi_{\pi}^{AV}(u),
\label{I1soft_pi}%
\end{equation}
where $\phi_{\pi}^{AV}(u)$ is the axial-vector (leading twist) pion
distribution amplitude:
\begin{equation}
\langle0|\bar{d}(z)\gamma_{\mu}\gamma_{5}u(-z)|\pi^{+}(p)\rangle=i\sqrt
{2}F_{\pi}p_{\mu}\int\limits_{0}^{1}du\,e^{i(2u-1)z\cdot p}\phi_{\pi}^{AV}(u).
\label{phiAV_me}%
\end{equation}
We shall come back to this point in Sect.\ref{softpions}.

\subsection{Pion skewed distributions}
\label{sect:SPD}

For $\pi$SPDs we have (for a review see Ref.\cite{RadAPPB}):
\begin{align}
&  \frac{1}{2}\int\frac{d\lambda}{2\pi}\exp\left(  i\lambda Xn\cdot\bar
{p}\right)  \left\langle \pi^{b}(p^{\prime})\left|  \bar{\psi}^{f^{\prime}%
}\left(  -\frac{\lambda n}{2}\right)  \rlap{/}n\psi^{f}\left(  \frac{\lambda
n}{2}\right)  \right|  \pi^{a}(p)\right\rangle \nonumber\label{spd_def}\\
&  =\delta^{ab}\,\delta^{ff^{\prime}}\,H^{I=0}(X,\xi,t)+i\varepsilon
^{abc}\,\left(  \tau^{c}\right)  ^{ff^{\prime}}\,H^{I=1}(X,\xi,t),
\end{align}
where $\bar{p}=(p+p^{\prime})/2$ is the average pion momentum.
SPDs depend on the following kinematical variables: the asymmetry
parameter $\xi=-\Delta ^{+}/(2\bar{p}^{+})$, where
$\Delta=p^{\prime}-p$ is the four-momentum transfer;
$t=\Delta^{2}$ and the variable $X$ defining the longitudinal
momenta of the struck and scattered quarks in DVCS,
$(X+\xi)\bar{p}^{+}$ and $(X-\xi)\bar{p}^{+}$ respectively.

The pion SPDs obey the following symmetry properties \cite{PW2}:
\begin{equation}
H^{I=0}(X,\xi,t)=H^{I=0}(X,-\xi,t)=-H^{I=0}(-X,\xi,t), \label{HI0_sym}%
\end{equation}%
\begin{equation}
H^{I=1}(X,\xi,t)=H^{I=1}(X,-\xi,t)=H^{I=1}(-X,\xi,t). \label{HI1_sym}%
\end{equation}
In the \textit{forward} limit the SPDs reduce to the usual quark and antiquark
distributions:
\begin{equation}
H^{I=0}(X,\xi=0,t=0)=\frac{1}{2}\left[  \Theta(X)q_{s}(X)-\Theta
(-X)q_{s}(-X)\right],
\label{HI0_fl}
\end{equation}
\begin{equation}
H^{I=1}(X,\xi=0,t=0)=\frac{1}{2}\left[  \Theta(X)q_{v}(X)+\Theta
(-X)q_{v}(-X)\right],
\label{HI1_fl}
\end{equation}
where $q_{s}(X)$, $q_{v}(X)$ are singlet (quark plus antiquark) and valence
(quark minus antiquark) distributions. For example for $\pi^{+}$ we have:
\begin{equation}
2H^{I=0}(X,\xi=0,t=0)=\left\{
\begin{array}{rcc}
u^{\pi^{+}}(X) + \bar{u}^{\pi^{+}}(X) & \mbox{for} & X > 0,\\
-d^{\pi^{+}}(-X) - \bar{d}^{\pi^{+}}(-X) & \mbox{for} & X < 0
\end{array}
\right.
\label{quarkH0}
\end{equation}
and
\begin{equation}
2H^{I=1}(X,\xi=0,t=0)=\left\{
\begin{array}{rcc}
u^{\pi^{+}}(X)-\bar{u}^{\pi^{+}}(X) & \mbox{for} & X > 0,\\
-d^{\pi^{+}}(-X)+\bar{d}^{\pi^{+}}(-X) & \mbox{for} & X < 0.
\end{array}
\right.   \label{quarkH1}%
\end{equation}
In the present model we get
$\bar{u}^{\pi^{+}}(X)=d^{\pi^{+}}(X)\equiv0$.

The normalization conditions for SPDs are analogous to the
$2\pi$DAs case (\ref{F_pi(s)}, \ref{r_pi_2piDA},
\ref{2piDAI0_norm}). In particular
\begin{equation}
\int\limits_{-1}^{+1}dX\,H^{I=1}(X,\xi,t)=F_{\pi}^{\mbox{{\small
em}}}(t), \label{F_pi(t)}%
\end{equation}%

\begin{figure}[h]
\begin{center}
\includegraphics[scale=1.2]{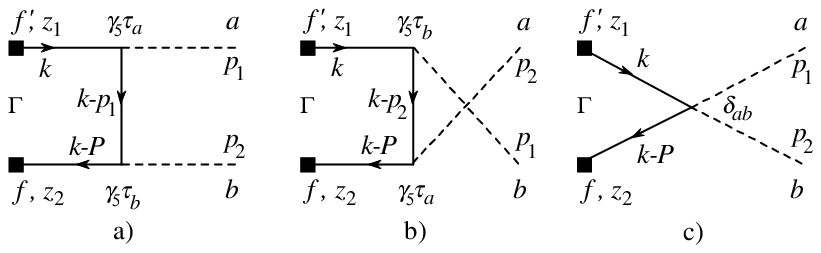}
\end{center}
\caption{{\footnotesize Diagrams contributing to the matrix
element
$\left\langle \pi^{a}(p_{1})\pi^{b}(p_{2})\left|  \bar{\psi}^{f^{\prime}%
}(z_{1})\Gamma\psi^{f}(z_{2})\right|  0\right\rangle $,
Eq.(\ref{2piDA_def}). Diagrams a) and b) contribute both to
isoscalar and isovector $2\pi$DAs;
diagram c) contributes only to the isoscalar $2\pi$DA. }}%
\label{2pi_diag}%
\end{figure}
%
%
%

\begin{equation}
\langle(r_{\pi}^{em})^{2}\rangle=6\left.  \frac{dF_{\pi}^{\mbox{{\small
em}}}(t)}{dt}\right|  _{t=0} \label{r_pi_spd}%
\end{equation}
and
\begin{equation}
\int\limits_{-1}^{+1}dX\,XH^{I=0}(X,\xi,t=0)=\frac{1}{2}(1-\xi^{2})M_{2}%
^{(\pi)}. \label{SPDI0_norm}%
\end{equation}
In analogy to the soft pion theorem for $2\pi$DAs, in the case of $\pi$SPD we
have \cite{P}:
\begin{equation}
H^{I=1}(X,\xi=1,t=0)=\phi_{\pi}^{AV}\left(  \frac{X+1}{2}\right)  ,\qquad
H^{I=0}(X,\xi=1,t=0)=0. \label{HI_xi1}%
\end{equation}


\section{Pion generalized distribution amplitudes in the effective model}
\label{model_GDA}
%
%

\subsection{Analytical results}
\label{Analytical}

For the matrix elements entering the definitions of $2\pi$DAs and
SPDs, Eqs.(\ref{2piDA_def},\ref{spd_def}), in the effective model
described in Sect.\ref{model}, we obtain:
\begin{align}
\lefteqn{\left\langle \pi^{a}(p_{1})\pi^{b}(p_{2})\left|
\bar{\psi}^{f\prime
}(z_{1})\Gamma\psi^{f}(z_{2})\right|0\right\rangle = i\frac{N_{C}}{F_{\pi}%
^{2}}\int\frac{d^{4}k}{(2\pi)^{4}}e^{ikz_{1}-i(k-P)z_{2}}}\nonumber\\
&  \left\{  \delta^{ab}\delta^{ff^{\prime}}\left[  \mathcal{T}_{1}%
(k-P,k)+\mathcal{T}_{2}\left(  k-P,k-p_{2},k\right)
+\mathcal{T}_{2}\left(
k-P,k-p_{1},k\right)  \right]  \right.  \nonumber\\
\qquad &  \left.
+i\varepsilon^{abc}\,(\tau^{c})^{ff^{\prime}}\left[
\mathcal{T}_{2}\left(  k-P,k-p_{2},k\right)
-\mathcal{T}_{2}\left(
k-P,k-p_{1},k\right)  \right]  \right\}  ,\label{2pi_model}%
\end{align}

%
%
\begin{figure}[h]
\begin{center}
\includegraphics[scale=1.2]{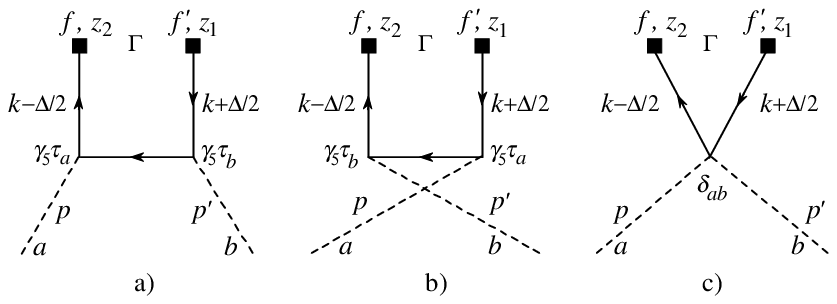}
\end{center}
\caption{{\footnotesize Diagrams contributing to the matrix
element
$\left\langle \pi^{b}(p^{\prime})\left|  \bar{\psi}^{f^{\prime}}(z_{1}%
)\Gamma\psi^{f}(z_{2})\right|  \pi^{a}(p)\right\rangle $, Eq.
(\ref{spd_def}). Diagrams a) and b) contribute both to isoscalar
and isovector SPDs; diagram c)
contributes only to the isoscalar SPD. }}%
\label{spd_diag}%
\end{figure}
%
%

\noindent and
\begin{align}
&  \left\langle \pi^{b}(p^{\prime})\left|  \bar{\psi}^{f^{\prime}}%
(z_{1})\Gamma\psi^{f}(z_{2})\right|  \pi^{a}(p)\right\rangle =i\frac{N_{C}%
}{F_{\pi}^{2}}\int\frac{d^{4}k}{(2\pi)^{4}}e^{i\left(  k+\frac{\Delta}%
{2}\right)  z_{1}-i\left(  k-\frac{\Delta}{2}\right)  z_{2}}\nonumber\\
&  \left\{  \delta^{ab}\delta^{ff^{\prime}}\left[
\mathcal{T}_{1}\left(
k-\frac{\Delta}{2},k+\frac{\Delta}{2}\right) \right. \right. \label{skewed_model}\\
&  \qquad\qquad\left.  +\mathcal{T}_{2}\left(
k-\frac{\Delta}{2},k-\bar
{p},k+\frac{\Delta}{2}\right)  +\mathcal{T}_{2}\left(  k-\frac{\Delta}%
{2},k+\bar{p},k+\frac{\Delta}{2}\right)  \right]  \nonumber\\
&  \left.  +i\varepsilon^{abc}\,\left(  \tau^{c}\right)
^{ff^{\prime}}\left[ \mathcal{T}_{2}\left(
k-\frac{\Delta}{2},k-\bar{p},k+\frac{\Delta}{2}\right)
-\mathcal{T}_{2}\left(  k-\frac{\Delta}{2},k+\bar{p},k+\frac{\Delta}%
{2}\right)  \right]  \right\},\nonumber %
\end{align}
where
\begin{equation}
\mathcal{T}_{1}(q,k)=\mbox{Tr}\left[  \Gamma\frac{1}{\rlap{/}q-M_{q}%
+i\epsilon}\sqrt{M_{q}M_{k}}\frac{1}{\rlap{/}k-M_{k}+i\epsilon}\right]
,\label{T1}%
\end{equation}%
\begin{equation}
\mathcal{T}_{2}(r,q,k)=\mbox{Tr}\left[
\Gamma\frac{\sqrt{M_{r}}}{\rlap
{/}r-M_{r}+i\epsilon}\gamma^{5}\frac{M_{q}}{\rlap{/}q-M_{q}+i\epsilon}%
\gamma^{5}\frac{\sqrt{M_{k}}}{\rlap{/}k-M_{k}+i\epsilon}\right]. \label{T2}%
\end{equation}
For the leading twist chirally even distributions
$\Gamma=\rlap{/}n$.

Equation (\ref{skewed_model}) is obtained, due to the crossing
symmetry, from (\ref{2pi_model}) by the exchange
$p_{1}\rightarrow-p$ and $p_{2}\rightarrow p^{\prime}$ (that is
$P\rightarrow\Delta$). The diagrams contributing to the matrix
elements (\ref{2pi_model}) and (\ref{skewed_model}) are shown in
figures \ref{2pi_diag} and \ref{spd_diag} respectively. The
diagrams a) and b) contribute both to the isoscalar and isovector
GDAs, while the diagram c) contributes only to the
isoscalar GDAs.

In the case of $2\pi$DAs we will work in the reference frame defined by
$\vec{P}_{\perp}= 0$. In this frame we find:
\begin{equation}
P = \left(  P^{+}, \, \frac{s} {P^{+}}, \, \vec{0}_{\bot} \right)  ,
\end{equation}%
\begin{equation}
p_{1} = \left(  v P^{+}, \, (1-v) \frac{s} {P^{+}}, \, \vec{p}_{\bot} \right)
, \qquad p_{2} = \left(  (1-v) P^{+}, \, v \frac{s} {P^{+}}, \, - \vec
{p}_{\bot}\right)  ,
\end{equation}%
\begin{equation}
p_{\bot}^{2} = v (1-v) s - m_{\pi}^{2}.
\end{equation}
From $p_{\bot}^{2}\geq0$ it follows
\begin{equation}
1 - \sqrt{1 - \frac{4m_{\pi}^{2}} {s}} \leq2 v \leq1 + \sqrt{1 - \frac
{4m_{\pi}^{2}} {s}}, \qquad s \geq4m_{\pi}^{2}.\label{v_limits}%
\end{equation}
In the case of pion SPDs we will work in the reference frame defined by
$\vec{\bar{p}}_{\perp}= 0$. In this frame we find:
\begin{align}
\bar{p}  &  = \left(  \bar{p}^{+}, \, \frac{-t + 4m_{\pi}^{2}} {4 \bar{p}^{+}%
}, \, \vec{0}_{\bot}\right)  = \left(  \bar{p}^{+}, \, \frac{-\tilde{t}} {4
\bar{p}^{+}}, \, \vec{0}_{\bot}\right)  ,\\
\Delta &  = \left(  -2 \xi\bar{p}^{+}, \, \xi\frac{-t + 4m_{\pi}^{2}} {2
\bar{p}^{+}} , \, \vec{\Delta}_{\bot} \right)  = \left(  -2 \xi\bar{p}^{+}, \,
\xi\frac{-\tilde{t}} {2 \bar{p}^{+}} , \, \vec{\Delta}_{\bot} \right)  ,\\
\Delta_{\bot}^{2}  &  = -t - \xi^{2} (-t + 4m_{\pi}^{2}) = (1- \xi^{2})
(-\tilde{t}) - 4m_{\pi}^{2}.
\end{align}
From $\Delta_{\bot}^{2}\geq0$ it follows
\begin{equation}
- \frac{\sqrt{-t}} {\sqrt{-t + 4m_{\pi}^{2}}} \leq\xi\leq\frac{\sqrt{-t}}
{\sqrt{-t + 4m_{\pi}^{2}}}.\label{xi_limits}%
\end{equation}
In what follows we take the chiral limit, $m_{\pi} = 0$. In this limit, from
(\ref{v_limits}) and (\ref{xi_limits}), we get $0 \leq v \leq1$ and $-1
\leq\xi\leq1$.

Inserting Eqs.(\ref{2pi_model}) and (\ref{skewed_model}) into
Eqs.(\ref{2piDA_def}) and (\ref{spd_def}) we find $k^{+}=uP^{+}$
and $k^{+}=X\bar{p}^{+}$ respectively.

The method of evaluating $dk^{-}$ integral, taking the full care
of the momentum mass dependence, has been given in \cite{PR}. To
evaluate $dk^{-}$ integral we have to find the poles in the
complex $k^{-}$ plane. It is important to note that the poles come
only from momentum dependence in the denominators of
Eqs.(\ref{T1},\ref{T2}). Indeed, for each quark line with the
momentum $k$, in the case $M_{k}\rightarrow\infty$ we have at most
$M_{k}$ in the numerator (if the line is coupled to pion lines at
both ends) and $M_{k}$ in the denominator. This means that the
position of the poles is given by the zeros of denominator, that
is by the solutions of the equation
\begin{equation}
k^{2}-M^{2}\left(  \frac{\Lambda^{2}}{k^{2}-\Lambda^{2}+i\epsilon}\right)
^{4n}+i\epsilon=0.
\end{equation}
This equation is equivalent to
\begin{equation}
G(z)=z^{4n+1}+z^{4n}-r^{2}=\prod_{i=1}^{4n+1}(z-z_{i}), \label{G}%
\end{equation}
with $z=k^{2}/\Lambda^{2}-1+i\epsilon$ and
$r^{2}=M^{2}/\Lambda^{2}$. For $r^{2}\neq0$ (or finite $\Lambda$)
equation (\ref{G}) has $4n+1$ nondegenerate solutions. In general
case $4n$ of them can be complex and the care must be taken about
the integration contour in the complex $k^{-}$ plane. Because of
the imaginary part of the $z_{i}$'s, the poles in the complex
$k^{-}$ plane can drift across Re$k^{-}$ axis. In this case the
contour has to be modified in such a way that the poles are not
allowed to cross it. This follows from the analyticity of the
integrals in the $\Lambda$ parameter and ensures the vanishing of
GDAs in the kinematically forbidden regions. We get nonvanishing
results for $2\pi$DAs and SPDs for $0\leq u\leq1$ and $-1\leq
X\leq1$ respectively.

%
%
\begin{figure}[h]
\begin{center}
$s=0$~GeV$^2$ \hspace{3.5cm} $s=1$~GeV$^2$\\
~~~~\\
\psfrag{u}{~} \psfrag{v}{~}
\psfrag{quark}{\footnotesize quark $u$} %
\psfrag{pion}{\footnotesize pion $v$}
\includegraphics[scale=0.5]{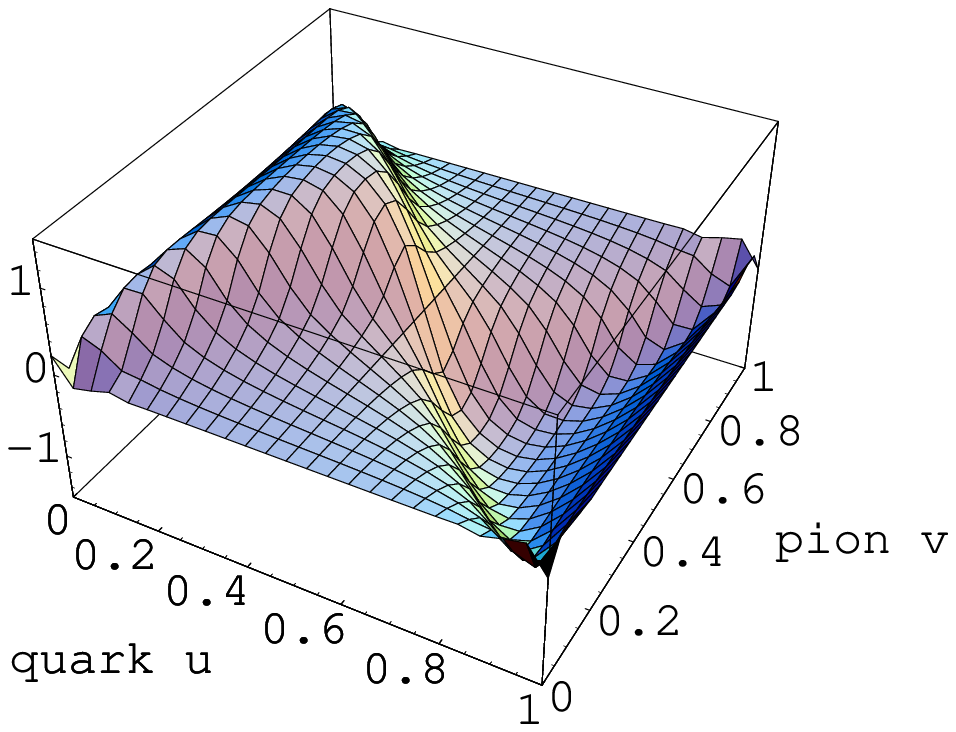}
\includegraphics[scale=0.5]{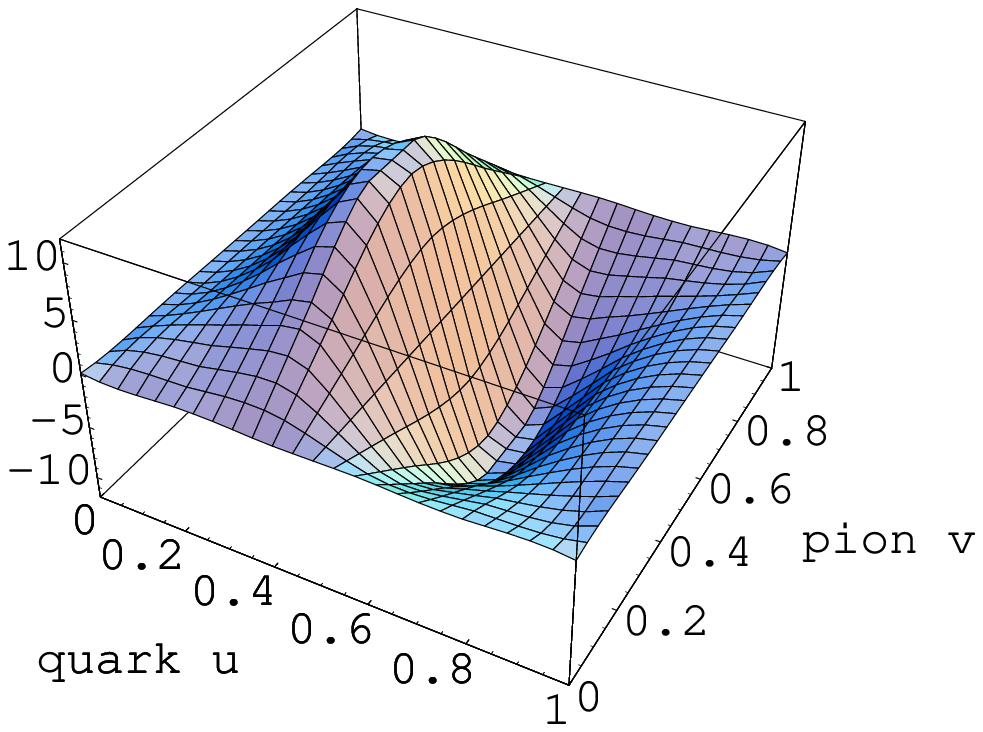} \\
\includegraphics[scale=0.5]{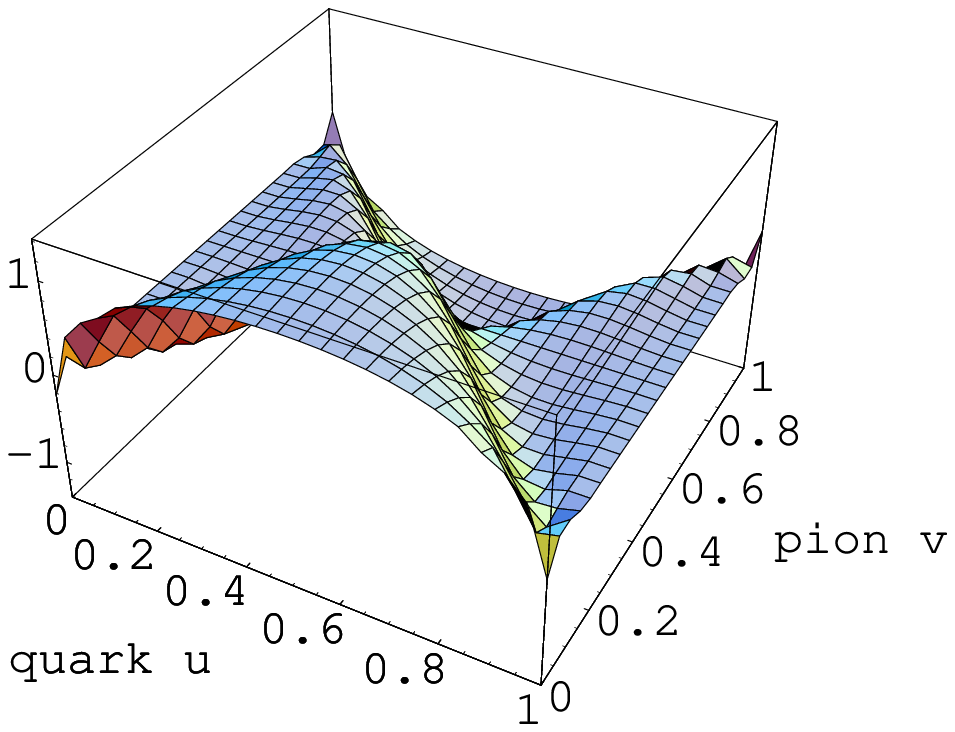}
\includegraphics[scale=0.5]{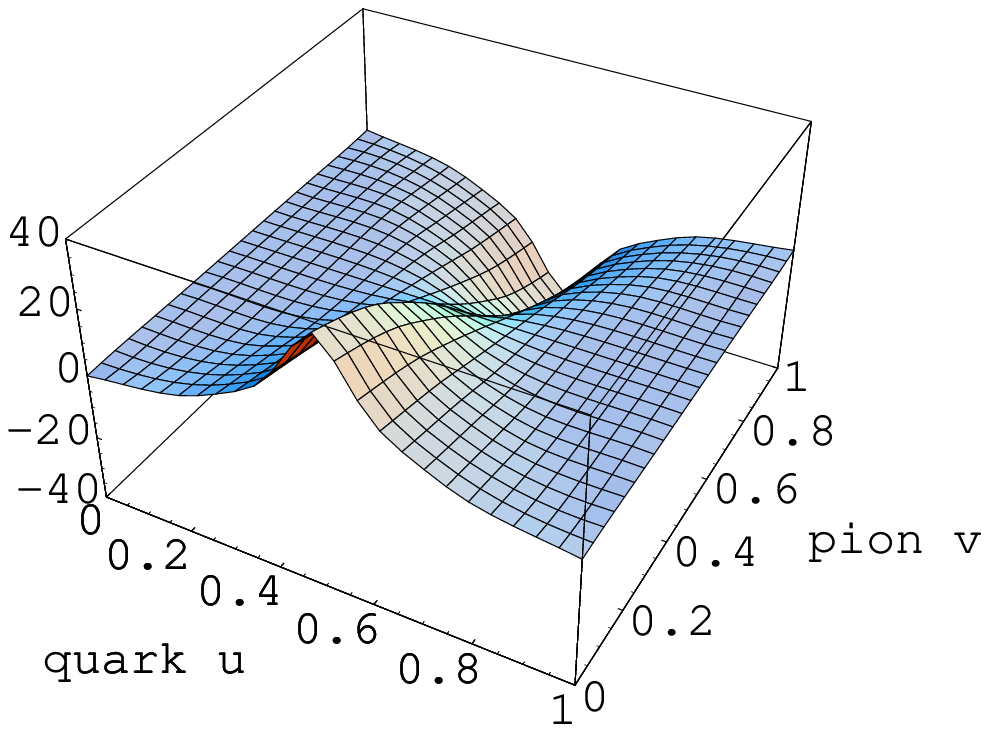}
\end{center}
\caption{{\footnotesize Chirally even isoscalar (upper row) and
isovector (lower row) 2$\pi$DA's for $M=350$~MeV and $n=1$ for two
different invariant masses of outgoing pions' momenta. Note that
for small $s$ and $v=0$  the isovector 2$\pi$DA resembles -- as
function of $u$ -- the axial-vector
one pion distribution amplitude in agreement with the soft pion theorem
(\ref{I1soft_pi},\ref{soft_even})}}%
\label{EDA}%
\end{figure}

After evaluating $dk^{-}$ integral the $d^{2}k_{\perp}$ integral
has to be treated numerically. The integral over angular
dependence in the transverse plane can be in principle evaluated
analytically (using residue technique), or in other words an exact
algorithm for its evaluation can be given. In this way we are left
with the numerical integration in only one variable, which is an
easy task to do. The technical details of the calculations are
presented in Appendix \ref{appA}.

%
%
\begin{figure}[h]
\begin{center}
$s=0$~GeV$^2$ \hspace{3.5cm} $s=1$~GeV$^2$\\
~~~~\\
\psfrag{u}{~} \psfrag{v}{~}
\psfrag{quark}{\footnotesize quark $u$} %
\psfrag{pion}{\footnotesize pion $v$}
\includegraphics[scale=0.5]{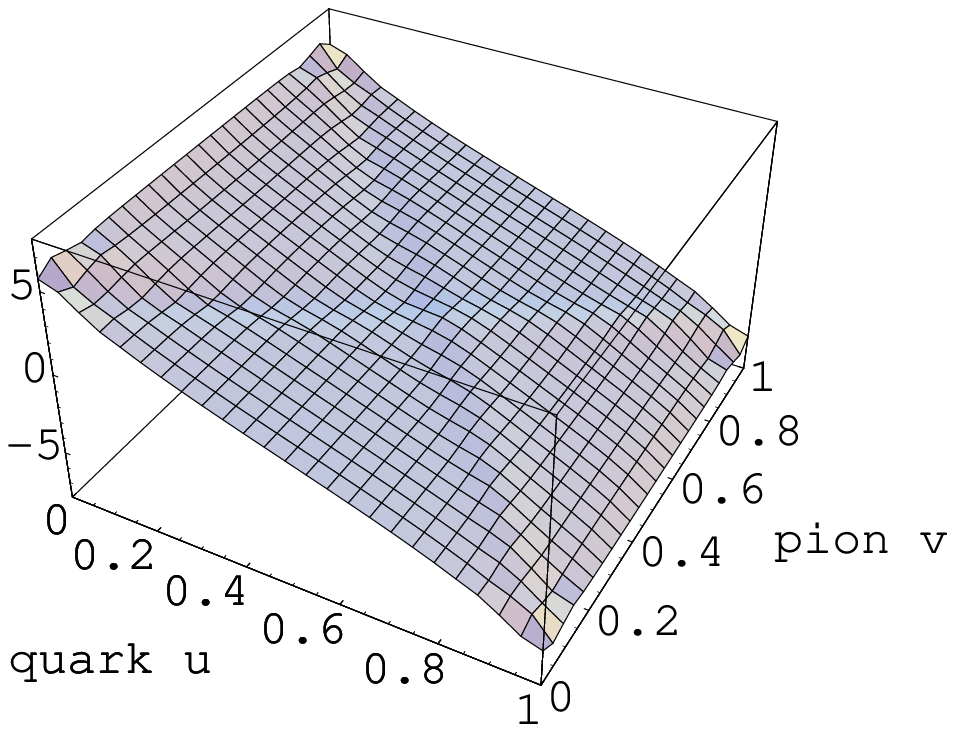}
\includegraphics[scale=0.5]{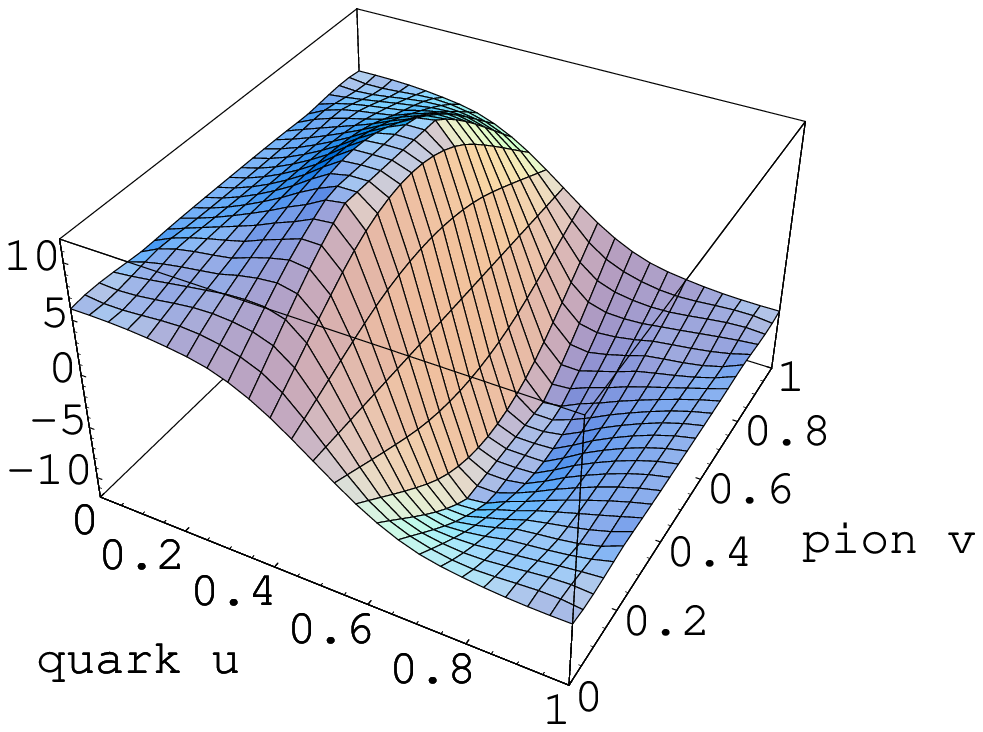}\\
\includegraphics[scale=0.5]{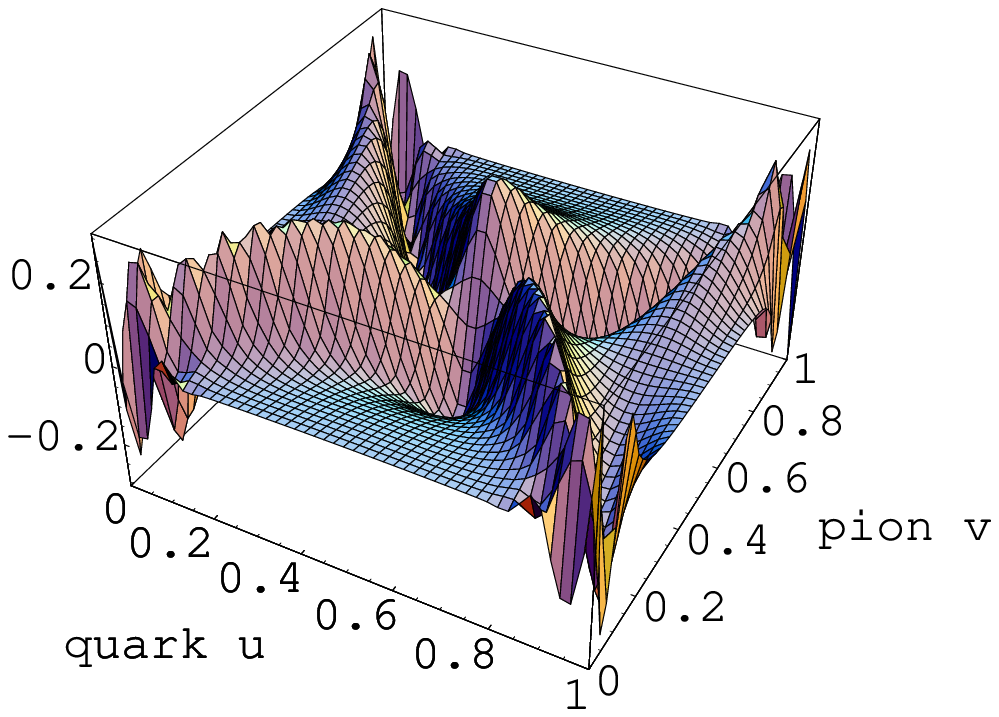}
\includegraphics[scale=0.5]{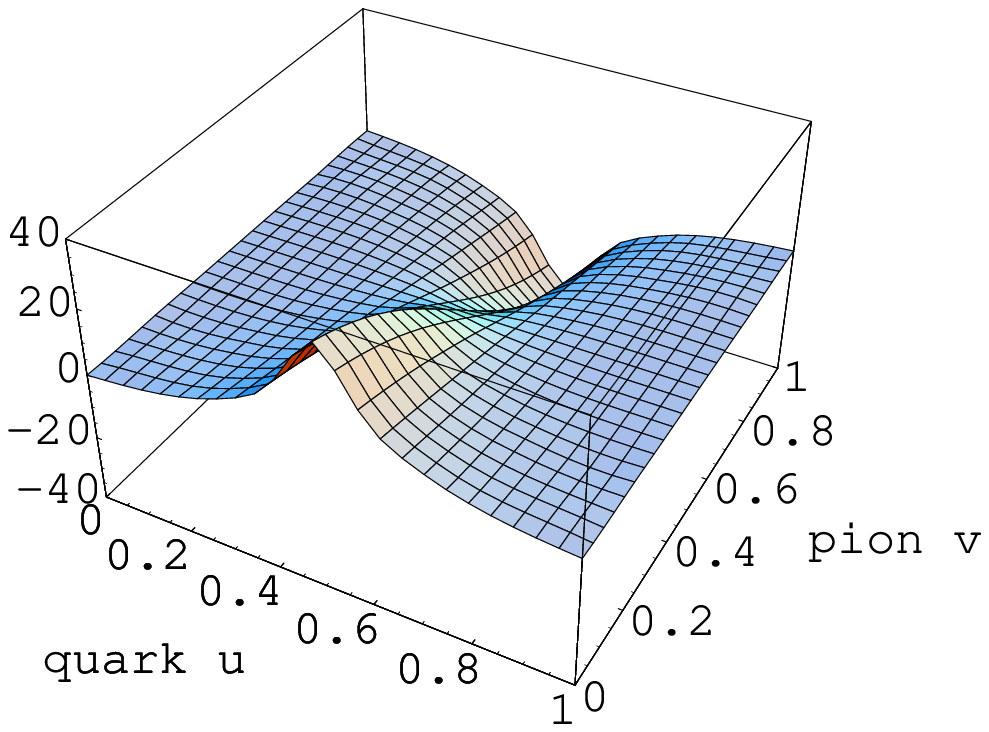}
\end{center}
\caption{{\footnotesize Chirally odd isoscalar (upper row) and
isovector (lower row) 2$\pi$DA's for $M=350$~MeV and $n=1$ for two
different invariant masses of outgoing pions' momenta. Note that
for small $s$ and $v=0$ the shape in $u$ of the isoscalar 2$\pi$DA
resembles the derivative (straight line) of the pseudo-tensor
one pion distribution amplitude in agreement with the soft pion theorem
(\ref{soft_odd}). }}%
\label{ODA}%
\end{figure}

\begin{figure}[h]
\begin{center}
\includegraphics[scale=0.6]{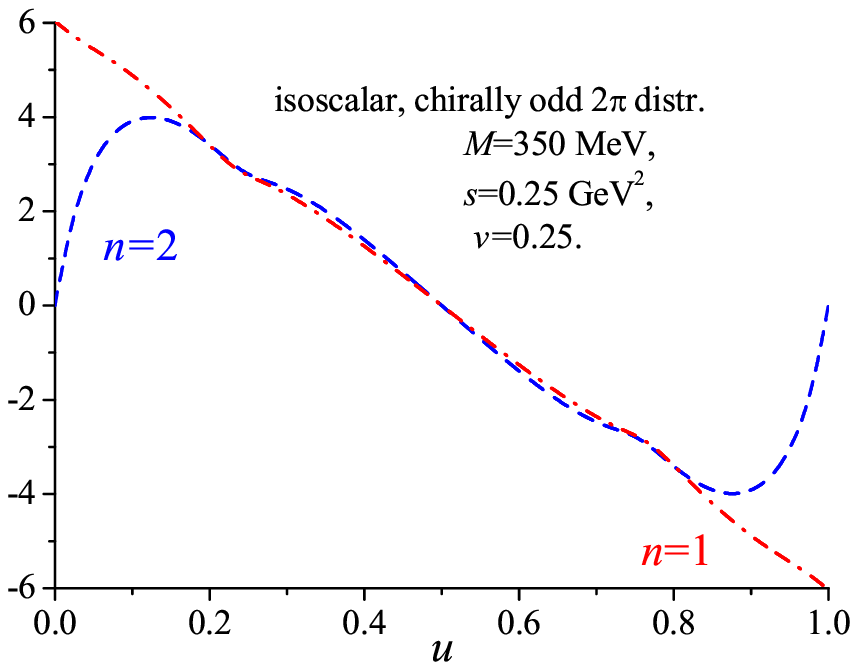}
\includegraphics[scale=0.6]{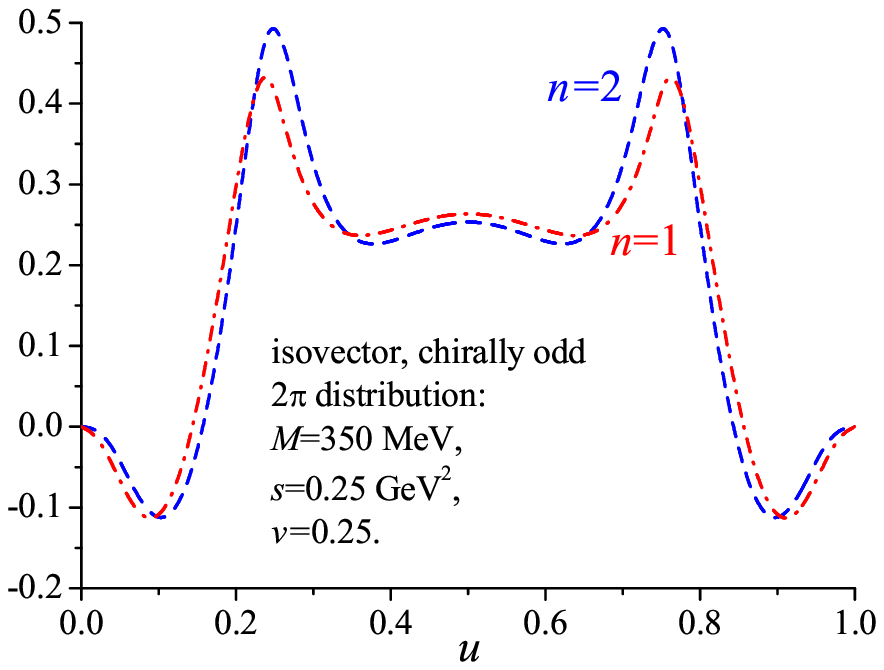}
\end{center}
\caption{{\footnotesize Isoscalar (left) and isovector (right)
chirally odd 2$\pi$DA in the chiral limit for $M=350$~MeV, $n=1$
(dashed-dotted) and $2$ (dashed line), $s=0.25$ GeV${}^2$, at $v=0.25$.}}
\label{ODAndep}%
\end{figure}

\subsection{Numerical results}
\label{Numerical}



\begin{figure}[h]
\begin{center}%
\psfrag{X}{\footnotesize $X$}
\psfrag{xi}{\footnotesize  $\xi$} %
\psfrag{t}{\footnotesize $t=0$~GeV$^2$}
\psfrag{t1}{\footnotesize$t=-1$~GeV$^2$}
\includegraphics[scale=0.5]{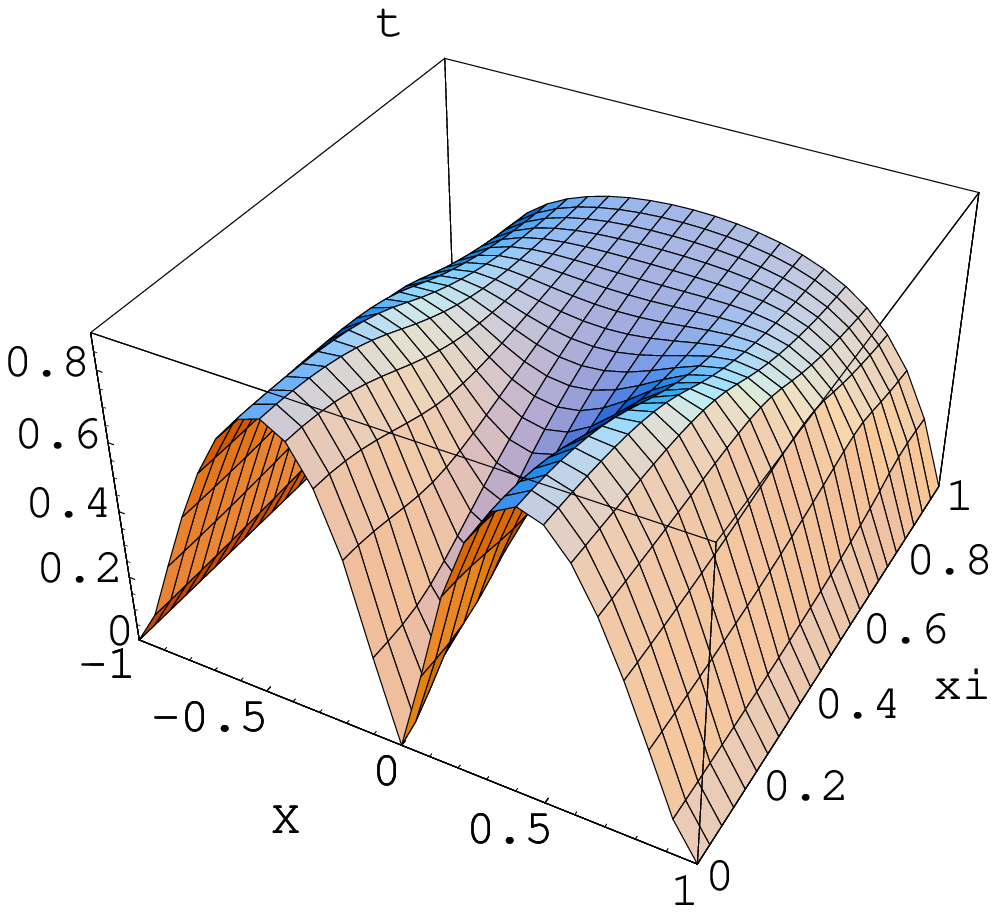}
\includegraphics[scale=0.5]{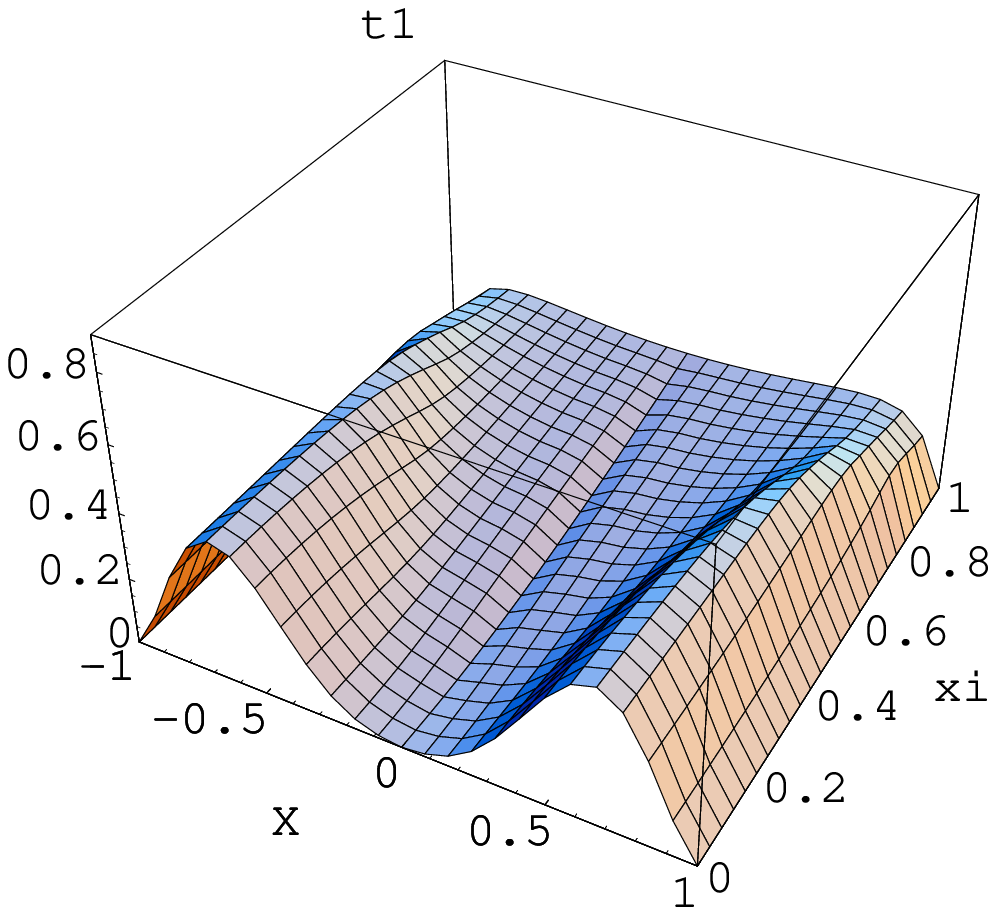} \\
\includegraphics[scale=0.5]{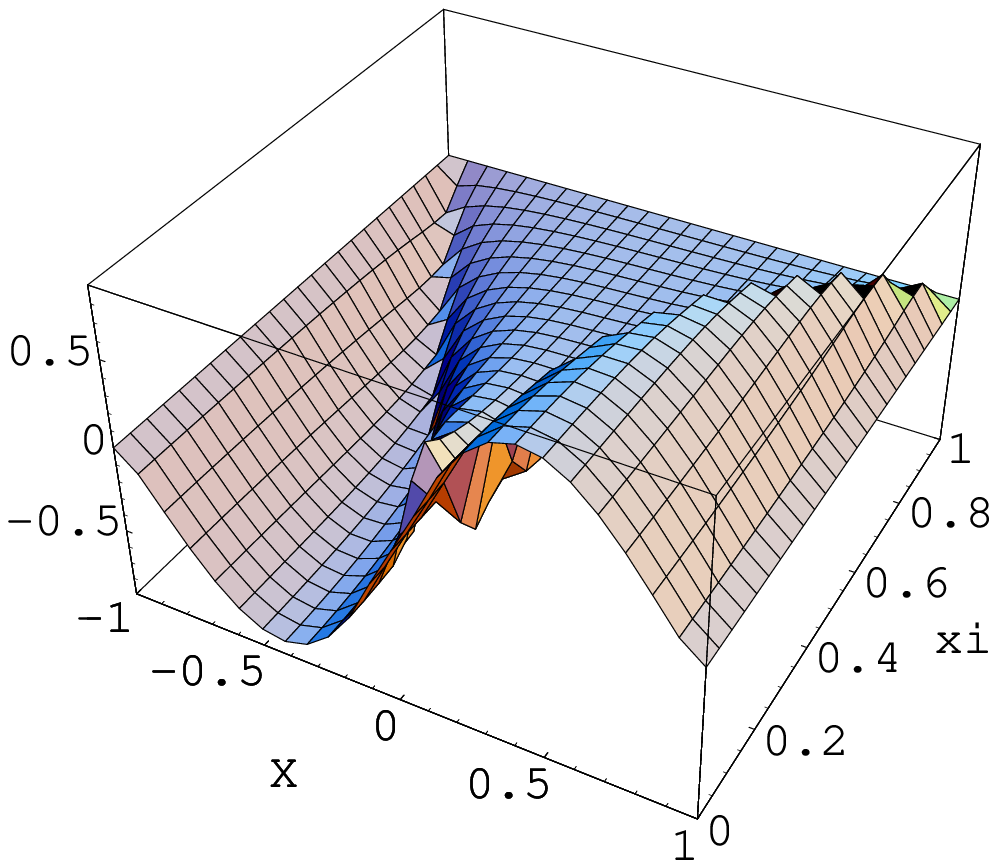}
\includegraphics[scale=0.5]{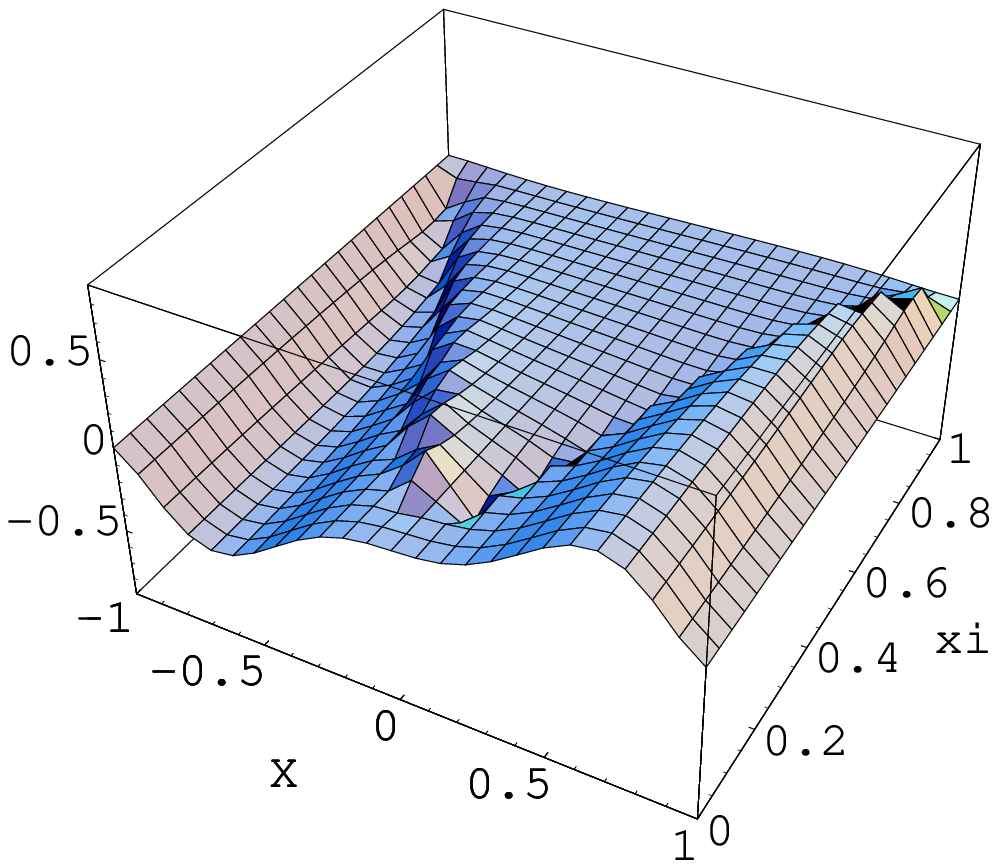}
\end{center}
\caption{{\footnotesize Isovector (upper row) and isoscalar (lower
row) pion SPD, $H^{I=1,0} (X, \xi, t)$, in the chiral
limit ($m_{\pi}=0$) for $M=350$ MeV; for $t=0$ and $-1$ GeV$^2$. }}%
\label{SPD}%
\end{figure}

Our results for $2\pi$DAs and SPDs are presented in Figs. \ref{EDA}%
-\ref{SPD}. Although strictly speaking the present model is valid only for
small pion momenta, $s$, we show also results for $s$ as large as 1 GeV in order
to trace the trends of the change of shape.

For small $s$ and one pion momentum equal zero ({\em i.e.} $v=0$)
the shape of the chirally even isovector 2$\pi$DA resembles -- as
function of $u$ -- the axial-vector one pion distribution
amplitude in agreement with the soft pion theorem
(\ref{I1soft_pi},\ref{soft_even}). Similarly, chirally odd
isoscalar 2$\pi$DA resembles the derivative (straight line) of the
pseudo-tensor one pion distribution amplitude, also in agreement
with the soft pion theorem (\ref{soft_odd}) which we shall discuss
in detail in Sect.{\ref{softpions}}. The functions that we obtain,
obey correct symmetry properties (\ref{I02piDA_sym},
\ref{I12piDA_sym}) and (\ref{HI0_sym}, \ref{HI1_sym}). The
asymptotic form for the isovector 2$\pi$DA is plotted in
Fig.\ref{asymp} and we see that it has a different shape from the
model prediction of Fig.\ref{EDA}.

As in the case of the pion distribution amplitude \cite{PR} the
$n$ dependence (see Eq.(\ref{M})) of our results is weak. This is
depicted in Fig.{\ref{ODAndep}} where we plot chirally odd
2$\pi$DA's for $M=350$~MeV and $n=1$ and 2, $s=0.25$ GeV${}^2$,
at $v=0.25$. One can see
that there is almost no $n$ dependence except for the end point
behavior of the isoscalar distribution which vanishes at $u=0,1$
for all $n >1 $. This kind of behavior has been already discussed
in Ref.\cite{PRcond} in the context of the pseudo-scalar one pion
DA. The $n$ dependence of the chirally even 2$\pi$DA is similarly
weak.

\begin{figure}[h]
\begin{center}%
\includegraphics[scale=0.85]{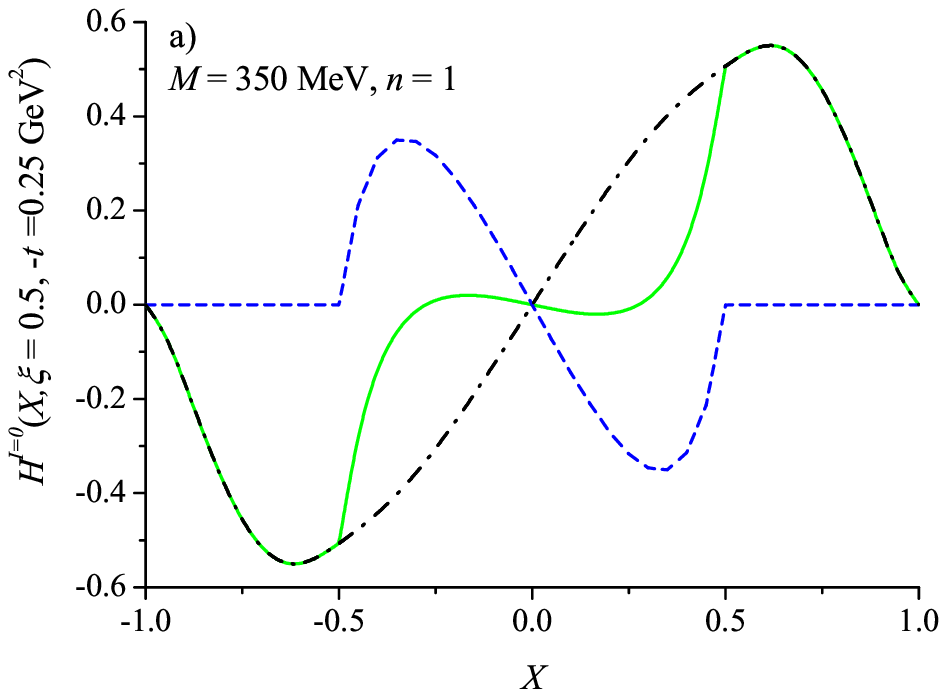}\\
\includegraphics[scale=0.85]{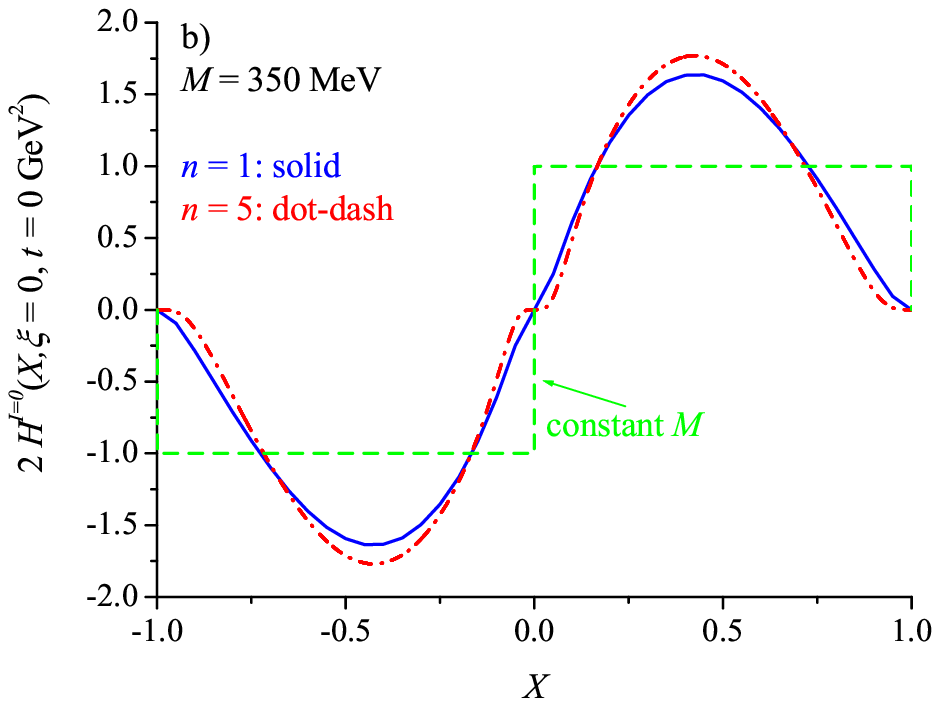}
\end{center}
\caption{{\footnotesize Isoscalar pion SPD, $H^{I=0} (X, \xi, t)$,
in the chiral limit ($m_{\pi}=0$) for $M=350$ MeV; for $t=-0.25$
and $0$ GeV$^2$ plotted as function of $X$ for fixed $\xi$. In
Fig.a) dashed-dotted curve corresponds to diagrams a)+b) of Fig.2, and
dashed line to diagram c), whereas the solid curve
corresponds to the sum. In the forward case (Fig.b)) we show
results for $n=1$ (solid) and 5 (dashed-dotted line), as well as
the distribution corresponding to constant $M(k)$ \cite{ERA}. Note
that in Fig.b.) $2H^{I=0}$ for $X<0$ is equal to $-\overline{d}(-X)$
distribution in $\pi^+$ while for $X>0$ it corresponds to $u(X)$.}}%
\label{figH}%
\end{figure}

Skewed pion distributions for $M=350$~MeV and $n=1$ are plotted in
Fig.\ref{SPD} for two values of $t=0$ and $-1$~GeV$^2$. Note that
we only plot them for $0<\xi<1$, the remainder in region of
$-1<\xi<0$ is symmetric according to
Eqs.(\ref{HI1_sym},\ref{HI0_sym}).
In the present model
\begin{equation}
H^{I=1} (X, \, \xi=0, \, t=0) =
\left\{
\begin{array}{rcc}
H^{I=0} (X, \, \xi=0, \, t=0) & \mbox{for} & X > 0 \\
-H^{I=0} (X, \, \xi=0, \, t=0) & \mbox{for} & X < 0
\end{array}
\right.
\end{equation}
and from (\ref{HI0_fl}, \ref{HI1_fl}) it follows that
$q_s(X)=q_v(X)$. The section of $H^{I=0}(X,\xi,t)$ as function of
$X$ is plotted in Fig.\ref{figH}. In Fig.\ref{figH}.a we plot
$H^{I=0}(X,\xi=0.5,t=-0.25~{\rm GeV}^2)$ for $M=350$~MeV and
$n=1$. The solid curve corresponds to the full result, whereas the
dashed-dotted curve corresponds to diagrams a)+b) of
Fig.\ref{spd_diag}, and dashed line to diagram c). In the forward
limit ($t=0$ and $\xi=0$, Fig.\ref{figH}.b) $H^{I=0}(X,\xi=0,t=0)$
corresponds to the quark densities in the pion (\ref{quarkH0}). We
see little $n$ dependence. For comparison we also plot quark
distributions obtained in the model with sharp cutoff in the
transverse momentum plane which coincides with the result obtained
in Refs.\cite{ERAPP,ERA}. These quark distributions are understood
to be at low normalization scale $Q_0$ and have to be evolved to
some physical scale $Q$ at which they are experimentally
accessible. This will be done in Sect.\ref{strucfun}.

Our skewed distributions do not exhibit factorization
\cite{RadFact} and for $\xi=0$ they are quite similar in shape to
the distributions obtained from local duality \cite{BRGS}.

%
%

\begin{figure}[b]
\begin{center}
\includegraphics[scale=0.85]{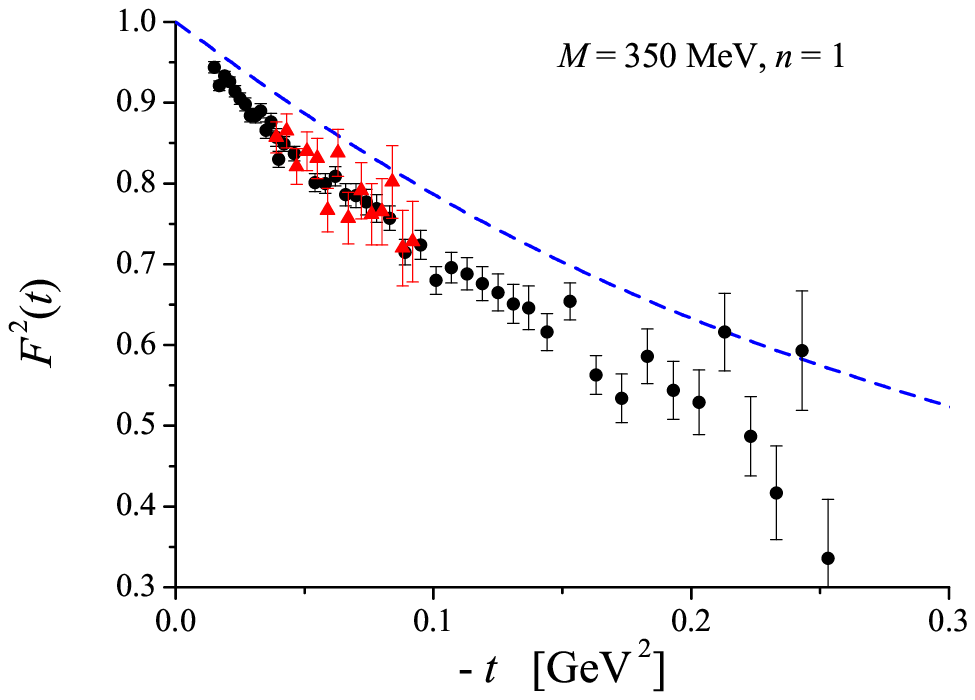} \\
\includegraphics[scale=0.85]{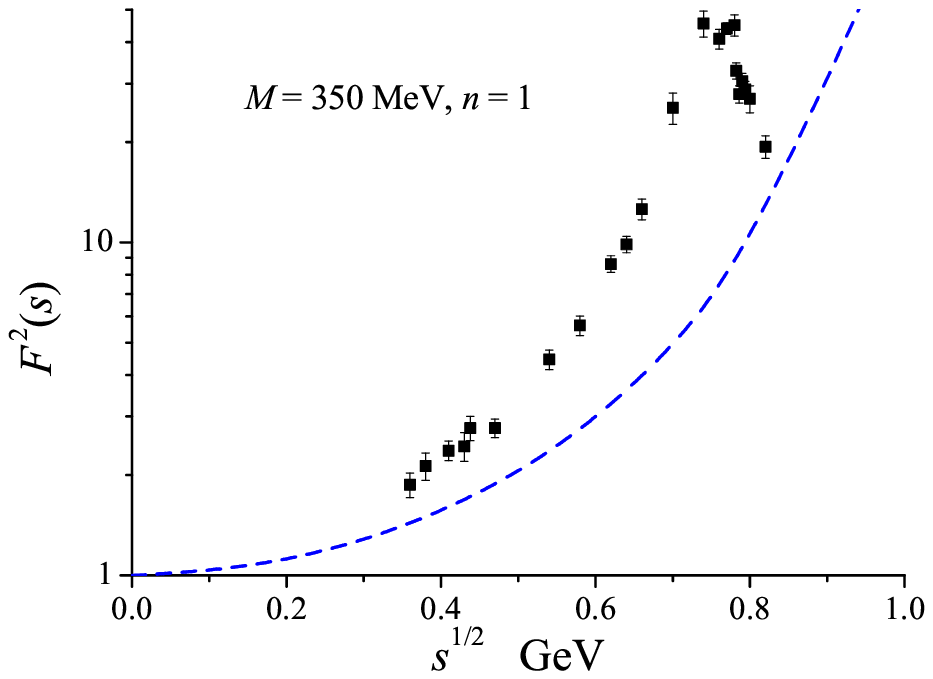}
\end{center}
\caption{{\footnotesize Pion form factors in space like (upper
fig.) and time like (lower fig.) regions. Theoretical curves
correspond to $M=350$~MeV and $n=1$. Experimental data are taken
from: dots \cite{Amendolia}, triangles \cite{Dally}, squares
\cite{timeff}.
}}%
\label{ffs}%
\end{figure}

The values of $F_{\pi}^{\mbox{{\small em}}}(s)$ and $F_{\pi}%
^{\mbox{{\small em}}}(t)$ evaluated by means of
Eqs.(\ref{F_pi(s)},\ref{F_pi(t)}) do not depend (as they should)
on $v$ and $\xi$ respectively. They are depicted in Fig.\ref{ffs},
together with the experimental data of Refs.
\cite{Amendolia}-\nocite{Dally}\cite{timeff}. We see that the space
like pion form factor overshoots the experimental points. This
discrepancy may be eventually cured by negative contribution from
the pion loops. In the time like region we see that very soon the
$\rho$ resonance tail switches on. Of course vector mesons are not
accounted for in the present model. The similar buildup of the
$\rho$ tail can be seen in the $\pi-\pi$ scattering data
\cite{MPGV}. One should note that strictly speaking our model can
be used only for very low momentum transfers. For higher momenta
QCD perturbative techniques should be used \cite{BRS}

The values of the pion electromagnetic radius, obtained from
Eqs.(\ref{r_pi_2piDA}) and (\ref{r_pi_spd}) are consistent with
each
other, although too small. Numerically we get $\langle(r_{\pi}^{em})^{2}%
\rangle=(0.54\mbox{
fm})^{2}$, which is comparable with the value
\[
\langle(r_{\pi}^{em})^{2}\rangle=\frac{N_{C}}{4\pi^{2}F_{\pi}^{2}}%
=(0.58\mbox{ fm})^{2},
\]

%
%

\noindent obtained within the instanton models. The experimental
value \cite{Amendolia} is, however, bigger:
$\langle(r_{\pi}^{em})^{2}\rangle=(0.66\mbox{ fm})^{2}$. This can
be, however, cured by the pion loops, which are neglected in the
present approach.

%
\section{Soft pion theorems}
\label{softpions}

Soft pion theorems relate 2$\pi$DAs for one pion momentum going to 0 with the
one pion light cone distribution amplitude.

We recall the definitions of the axial-vector and pseudo-tensor one pion DAs:
\begin{equation}
\phi_{\pi}^{AV}(u)=\frac{1}{i\sqrt{2}F_{\pi}}\int\limits_{-\infty}^{+\infty
}\frac{d\lambda}{\pi}e^{-i\lambda(2u-1)(n\cdot P)}\left\langle 0\right|
\bar{d} (\lambda n)\,\rlap{/}n\gamma_{5}\,u(-\lambda n)\left|
\pi^{+}(P)\right\rangle
\label{AV_phi}
\end{equation}
\begin{align}
\frac{d}{du}\phi_{\pi}^{PT}(u) = & -\frac{6F_{\pi}}{i\sqrt{2}\left\langle \bar
{q}q\right\rangle }\int\limits_{-\infty}^{+\infty}\frac{d\lambda}{\pi
}e^{-i\lambda(2u-1)(n\cdot P)}
\nonumber\\
& \times \left\langle 0\right|  \bar{d}(\lambda
n)\,in^{\alpha}P^{\beta}\sigma_{\alpha\beta}\gamma_{5}\,u(-\lambda
n)\left|  \pi^{+}(P)\right\rangle .\label{PT_phi}%
\end{align}
The matrix element entering the definitions (\ref{AV_phi}, \ref{PT_phi}),
calculated in the present model, gives:
\begin{equation}
\left\langle 0\right|  \bar{d}(z_{1})\gamma u(z_{2})\left|
\pi^{+}(P)\right\rangle  = -\frac{\sqrt{2}N_{C}}
{F_{\pi}}\int\frac{d^{4}k}{(2\pi)^{4}}e^{i(k-P)z_{1}%
-ikz_{2}}\,\mathcal{T}(k,k-P),
\label{onepi}
\end{equation}
with
\begin{align}
\mathcal{T}(k,k-P) &  =\mbox{Tr}\left(  \gamma\frac{\sqrt{M_{k}}}{\rlap
{/}k-M_{k}+i\epsilon}\gamma^{5}\frac{\sqrt{M_{k-P}}}{(\rlap{/}k-\rlap
{/}P)-M_{k-P}+i\epsilon}\right)  \nonumber\\
&  =\frac{\sqrt{M_{k}M_{k-P}}\,\mbox{Tr}\left(  \gamma\gamma^{5}[-\rlap
{/}k+M_{k}][(\rlap{/}k-\rlap{/}P)+M_{k-P}]\right)  }{\left[  k^{2}-M_{k}%
^{2}+i\epsilon\right]  \left[  (k-P)^{2}-M_{k-P}^{2}+i\epsilon\right]
}\label{T}%
\end{align}
Further calculations require an explicit form of $\gamma$.
Substituting (\ref{onepi}) into (\ref{AV_phi}) and (\ref{PT_phi}) we get
\begin{equation}
\phi_{\pi}^{AV}(u)=\frac{iN_{C}}{F_{\pi}^{2}}\int\frac{d^{4}k}{(2\pi)^{4}%
}\delta(n\cdot(k-u\,P))\,\mathcal{T}(k,k-P)|_{\gamma=\rlap{/}n\gamma_{5}}%
\end{equation}
and
\begin{equation}
\frac{d}{du}\phi(u)=-\frac{6iN_{C}}{\left\langle \bar{q}q\right\rangle }%
\int\frac{d^{4}k}{(2\pi)^{4}}\delta(n\cdot(k-u\,P))\,\mathcal{T}%
(k,k-P)|_{\gamma=in^{\alpha}P^{\beta}\sigma_{\alpha\beta}\gamma_{5}}
\end{equation}
respectively. Similarly, in the case of $2\pi$DAs, from (\ref{2piDA_def})
and (\ref{2pi_model}) we get
\begin{align}
\Phi_{2\pi}^{I=0}(u,v,s) = & \frac{iN_{C}}
{F_{\pi}^{2}}\int\frac{d^{4} k}{(2\pi)^{4}}\delta(n\cdot(k-u\,P))
\nonumber\\
& \left[  \mathcal{T}_{1}(k-P,k)+\mathcal{T}_{2}\left(  k-P,k-p_{2},k\right)
+\mathcal{T}_{2}\left(  k-P,k-p_{1},k\right)  \right]
\label{2pi_isoscalar}
\end{align}
and
\begin{align}
\Phi_{2\pi}^{I=1}(u,v,s) = &
\frac{iN_{C}}{F_{\pi}^{2}}\int\frac{d^{4}k}{(2\pi)^{4}}\delta(n\cdot(k-u\,P))
\nonumber\\
& \left[  \mathcal{T}_{2}\left(k-P,k-p_{2},k\right)
-\mathcal{T}_{2}\left(  k-P,k-p_{1},k\right)  \right],
\label{2pi_isovector}
\end{align}
with $\mathcal{T}_{1}$ and $\mathcal{T}_{2}$ defined in (\ref{T1}) and (\ref{T2}).
Soft pion theorems relate matrix elements (\ref{onepi}) and (\ref{2pi_model})
for one pion momentum, say $p_{2}\rightarrow0$ and for different $\gamma$ and
$\Gamma$. For $\mathcal{T}_{1}$ we have:
\begin{equation}
\mathcal{T}_{1}(k-P,k)=\frac{\sqrt{M_{k-P}M_{k}}\;
\mbox{Tr}\left(\Gamma\left[  (\rlap{/}k-\rlap{/}P)+M_{k-P}\right]
\left[  \rlap{/}k+M_{k}\right]  \right)  } {\left[
(k-P)^{2}-M_{k-P}^{2}+i\epsilon\right] \left[
k^{2}-M_{k}^{2}+i\epsilon\right]  }, \label{T1_1}
\end{equation}
whereas for the two $\mathcal{T}_{2}$'s in the limit $p_{2}%
\rightarrow0$ (that is $v\rightarrow1$) we get:%
\begin{align}
\mathcal{T}_{2}(k-P,k,k)  & =-\frac{\sqrt{M_{k-P}M_{k}}M_{k}\;\mbox{Tr}\left(
\Gamma\left[  \left(  \rlap{/}k-\rlap{/}P\right)  +M_{k-P}\right]  \right)
}{\left[  (k-P)^{2}-M_{k-P}^{2}+i\epsilon\right]  \left[  k^{2}-M_{k}%
^{2}+i\epsilon\right]  },\label{T2_a}\\
\mathcal{T}_{2}(k-P,k-P,k)  & =-\frac{\sqrt{M_{k-P}M_{k}}M_{k-P}\;\mbox
{Tr}\left(  \Gamma\left[  \rlap{/}k+M_{k}\right]  \right)  }{\left[
(k-P)^{2}-M_{k-P}^{2}+i\epsilon\right]  \left[  k^{2}-M_{k}^{2}+i\epsilon
\right]  }.\label{T2_b}%
\end{align}

\subsection{Chirally even $2\pi$DA and axial-vector wave function}

Consider first the soft pion theorem discussed already at the end
of Sect.\ref{sect:2pi}. To this end let us take%
\begin{equation}
\gamma=\rlap{/}n\gamma_{5}.
\end{equation}
Then
\begin{equation}
\mathcal{T}(k,k-P)|_{\Gamma=\rlap{/}n\gamma_{5}}=\frac{\sqrt{M_{k}M_{k-P}%
}\left[  -M_{k-P}\,\mbox{Tr}\left(  \rlap{/}n\rlap{/}k\right)  +M_{k}%
\;\mbox{Tr}\,\left(  \rlap{/}n(\rlap{/}k-\rlap{/}P)\right)  \right]  }{\left[
k^{2}-M_{k}^{2}+i\epsilon\right]  \left[  (k-P)^{2}-M_{k-P}^{2}+i\epsilon
\right]  }.
\end{equation}
For two pions we have to take
\begin{equation}
\Gamma=\rlap{/}n.
\end{equation}
Then in the soft limit $p_{1}=P$, $p_{2}=0$ (this means $v=1$) the isovector
combination in Eq.(\ref{2pi_isovector}) reads
\begin{align}
&  \mathcal{T}_{2}(k-P,k,k)|_{\Gamma=\rlap{/}n}-\mathcal{T}_{2}%
(k-P,k-P,k)|_{\Gamma=\rlap{/}n}\nonumber\\
\;\;\;\; &  =-\frac{\sqrt{M_{k-P}M_{k}}\,\left[  M_{k}\mbox{Tr}(\rlap
{/}n(\rlap{/}k-\rlap{/}P))-M_{k-P}\mbox{Tr}(\rlap{/}n\rlap{/}k)\right]
}{\left[  (k-P)^{2}-M_{k-P}^{2}+i\epsilon\right]  \left[  k^{2}-M_{k}%
^{2}+i\epsilon\right]  }\nonumber\\
&  =-\mathcal{T}(k-P,k)|_{\gamma=\rlap{/}n\gamma_{5}}%
\end{align}
Similarly, for the sum entering the isoscalar combination in
Eq.(\ref{2pi_model}) we have:%
\begin{equation}
\mathcal{T}_{1}(k-P,k)|_{\Gamma=\rlap{/}n}+\mathcal{T}_{2}(k-P,k,k)|_{\Gamma
=\rlap{/}n}+\mathcal{T}_{2}(k-P,k-P,k)|_{\Gamma=\rlap{/}n}=0.
\end{equation}
Hence the soft pion theorem for the chirally even $2\pi$DA takes the following form%
\begin{equation}
\Phi_{2\pi,\,\chi even}^{I=0}(u,v=1,s=0)=0,
\end{equation}%
\begin{equation}
\Phi_{2\pi,\,\chi even}^{I=1}(u,v=1,s=0)=-\phi_{\pi}^{AV}(u).
\label{soft_even}
\end{equation}

\subsection{Chirally odd $2\pi$DA and pseudo-tensor wave function}

Let us now consider chirally odd $2\pi$DA in the soft pion limit. To this end
let us take%
\begin{equation}
\gamma=in^{\alpha}P^{\beta}\sigma_{\alpha\beta}\gamma_{5}=\frac{1}{2}\left[
\rlap{/}P,\rlap{/}n\right]  \gamma_{5}%
\end{equation}
whose matrix element defines the derivative of pseudo-tensor one pion
distribution amplitude. Then%
\begin{equation}
\mathcal{T}(k,k-P)|_{\gamma=\frac{1}{2}\left[  \rlap{/}P,\rlap{/}n\right]
\gamma_{5}}=-\frac{1}{2}\frac{\sqrt{M_{k}M_{k-P}}\,\mbox{Tr}\left[  \rlap
{/}P,\rlap{/}n\right]  \rlap{/}k(\rlap{/}k-\rlap{/}P)}{\left[  k^{2}-M_{k}%
^{2}+i\epsilon\right]  \left[  (k-P)^{2}-M_{k-P}^{2}+i\epsilon\right]
}.\label{T_PT}%
\end{equation}
For 2 pions let us take%
\begin{equation}
\Gamma=in^{\alpha}P^{\beta}\sigma_{\alpha\beta}=\frac{1}{2}\left[  \rlap
{/}P,\rlap{/}n\right]  .
\end{equation}
Here only $\mathcal{T}_{1}$ survives%
\begin{align}
\mathcal{T}_{1}(k-P,k)|_{\Gamma=\frac{1}{2}\left[
\rlap{/}P,\rlap{/}n\right] } &
=\frac{1}{2}\frac{\sqrt{M_{k-P}M_{k}}\mbox{Tr}\left(  \left[ \rlap
{/}P,\rlap{/}n\right]  (\rlap{/}k-\rlap{/}P)\rlap{/}k\right)
}{\left[ (k-P)^{2}-M_{k-P}^{2}+i\epsilon\right]  \left[
k^{2}-M_{k}^{2}+i\epsilon
\right]  }\nonumber\\
&  =\mathcal{T}(k,k-P)|_{\gamma=\frac{1}{2}\left[  \rlap{/}P,\rlap{/}n\right]
\gamma_{5}}%
\end{align}
while $\mathcal{T}_{2}=0$. Therefore the soft pion theorem relates
the isoscalar part of the chirally odd $2\pi$DA to the derivative
of pseudo-tensor one pion amplitude:%
\begin{equation}
\Phi_{2\pi,\,\chi odd}^{I=0}(u,v=1,s=0)=-\frac{\left\langle \bar
{q}q\right\rangle }{6 F_{\pi}^{2}}\frac{d}{du}\phi_{\pi}%
^{PT}(u), \label{soft_odd}
\end{equation}%
\begin{equation}
\Phi_{2\pi,\,\chi odd}^{I=1}(u,v=1,s=0)=0
\end{equation}
This result cannot be derived by current algebra. Since the pseudo-tensor one
pion distribution amplitude is close to the asymptotic form $6u(1-u)$
\cite{PRcond} therefore we expect $\Phi_{2\pi,\chi odd}^{I=0}(u,v=0,s=0)\sim
1-2u$. That this is indeed the case can be seen from Fig.\ref{ODAndep}.a.
We recall that with our definitions (\ref{2piDA_def}) and
(\ref{Gamma_odd}) the chirally odd $2\pi$DAs have the dimension of mass.

%
%

\section{Pion structure function}
\label{strucfun}

As explained in Sect.\ref{sect:2pi} skewed pion distributions are
in the forward limit directly related to the parton distributions
inside the pion. For example, as explained after
Eqs.(\ref{quarkH0},\ref{quarkH1}) and as seen from
Fig.\ref{figH}.b
\begin{equation}
u^{\pi^{+}}(x)=\bar{d}^{\pi^{+}}(x)=2H^{I=0}(x,\xi=0,t=0).
\label{q_H_def}%
\end{equation}
On the other hand one can relate pion quark distributions to the
wave functions for all Fock states and polarizations
\cite{BrodskyLepage}. For $\pi^{+}$ we have
\begin{eqnarray}
u^{\pi^{+}}(x)&=&\sum\limits_{\mbox{Fock
states}}\sum\limits_{\lambda}\int
\frac{d^{2}k_{T}}{2\left(  2\pi\right)  ^{3}}\,\left|  \psi_{\lambda}%
(x,k_{T})\right|  ^{2}\nonumber \\
 &= & \int\frac{d^{2}k_{T}}{2\left(  2\pi\right)  ^{3}%
}\,\left|  \psi^{AV}(x,k_{T})\right|  ^{2}+\ldots\label{u_psi}%
\end{eqnarray}
where we have saturated the sum over the different Fock components
and polarizations by the axial-vector wave function. The function
$\psi^{AV}$ is (for $N_{c}=3$) normalized in the following way
\cite{BrodskyLepage}:
\begin{equation}
\int\limits_{0}^{1}dx\,\int\frac{d^{2}k_{T}}{\left(  2\pi\right)  ^{3}}%
\psi^{AV}(x,\vec{k}_{T})=\frac{F_{\pi}}{\sqrt{3}}. \label{norma}%
\end{equation}

It is clear that the normalization condition for $u^{\pi^{+}}$%
\begin{equation}
\int\limits_{0}^{1}dx\,u^{\pi^{+}}(x)=1 \label{unorma}%
\end{equation}
and normalization condition (\ref{norma}) are in general
incompatible since other components denoted by dots in
Eq.(\ref{u_psi}) are of importance. For example in a model with a
constant $M$ one obtains for $\psi^{AV}$ \cite{PP}:
\begin{equation}
\psi(u,\vec{k}^{AV}_{T})=\Theta\left(  u(1-u)\right)
\frac{2\sqrt{3}}{F_{\pi}%
}\frac{M^{2}}{\vec{k}_{T}^{2}+M^{2}}%
\end{equation}
and the normalization condition (\ref{norma}) is achieved by imposing a cutoff
$\Lambda$ on the transverse momentum integration $dk^2_{T}$:%
\begin{equation}
F_{\pi}^{2}=\frac{N_{c}M^{2}}{\left(  2\pi\right)  ^{2}}\ln\frac{M^{2}%
+\Lambda^{2}}{M^{2}}.
\end{equation}
It is now straightforward to calculate $u^{\pi^{+}}$%
\begin{equation}
u^{\pi^{+}}(x)=\Theta\left(  x(1-x)\right)  \frac{6}{\left(  2\pi\right)
^{2}}\frac{M^{2}}{F_{\pi}^{2}}\left[  1-\exp\left(  -\frac{\left(
2\pi\right)  ^{2}F_{\pi}^{2}}{N_{c}M^{2}}\right)  \right]  .
\end{equation}
The normalization constant is, as expected, smaller than 1,
however, for $M=350$~MeV we get 0.9 -- a fairly satisfactory
result for such a simplistic model. Of course the properly defined
quark distribution is also properly normalized \cite{ERAPP,ERA}.

\begin{figure}[h]
\begin{center}
\includegraphics[scale=0.7]{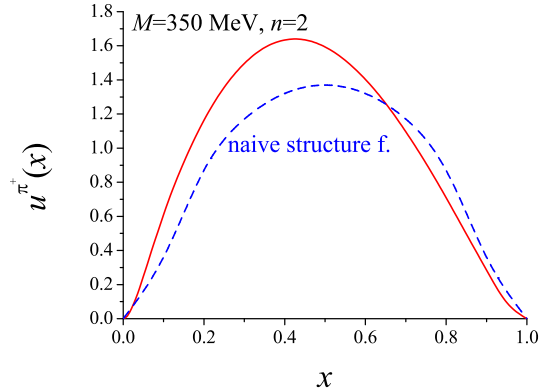}
\end{center}
\caption{{\footnotesize Comparison of the $u$ quark distribution
in $\pi^{+}$ defined
by Eq.(\ref{q_H_def}) (solid line) and the ''naive'' one defined by
Eq.(\ref{u_psi}) (dashed line) for $M=350$ MeV and $n=2$.}}%
\label{naive}%
\end{figure}

In the present model \cite{PR}
\begin{equation}
\psi^{AV}(x,\vec{k}_T)=\frac{2\sqrt{3}M^{2}}{\Lambda^{2}F_{\pi}}\sum\limits_{i,k}%
f_{i}f_{k}\,\frac{z_{i}^{n}z_{k}^{3n}x+z_{i}^{3n}z_{k}^{n}(1-x)}%
{\frac{\displaystyle \vec{k}_T^2}{\displaystyle \Lambda^{2}}+1+z_{i}x+z_{k}(1-x)}%
\end{equation}
and the normalization condition for $M=350$~MeV and $n=2$ gives
$0.88$ instead of $1$. The shape is also different from the
properly normalized result obtained with the help of
Eq.(\ref{q_H_def}), however, it can be explicitly seen that for
large $x$ both definitions converge, as they should \cite{JKR}.
This is depicted in Fig.\ref{naive}.
%
%
%
%

One of the major problems of the effective models like the one
considered here, is the normalization scale $Q_{0}$ at which the
model is defined. This is crucial shortcoming as far as the
comparison with the experimental data is concerned. It is argued
that the relevant scale for the instanton motivated models is of
the order of the inverse instanton size $1/\rho$ \emph{i.e.}
approximately 600 MeV. The precise definition of $Q_{0}$ is only
possible within QCD and in all effective models one can use only
qualitative \emph{order of magnitude} arguments to estimate
$Q_{0}$. A more practical way to determine $Q_{0}$ was discussed
in Ref.\cite{PRcond} where we associated $Q_{0}$ with the
transverse integration cutoff $K_{T}$ which was of the order of
$760<K_{T}<1100$~MeV.

On the other hand in Refs.\cite{ERA} it was argued that the
pertinent model scale may be as small as 350 MeV. This estimate is
based on the requirement that the total momentum carried by the
quarks equals to the one measured experimentally at
$Q=4$~GeV$^{2}$, \emph{i.e.} 47\% . In that way the initial
evolution scale $Q_{0}$ can be adjusted. It is, however,
problematic whether one can use QCD evolution equations at such
low $Q_{0}$.

\begin{figure}[h]
\begin{center}%
\includegraphics[scale=0.8]{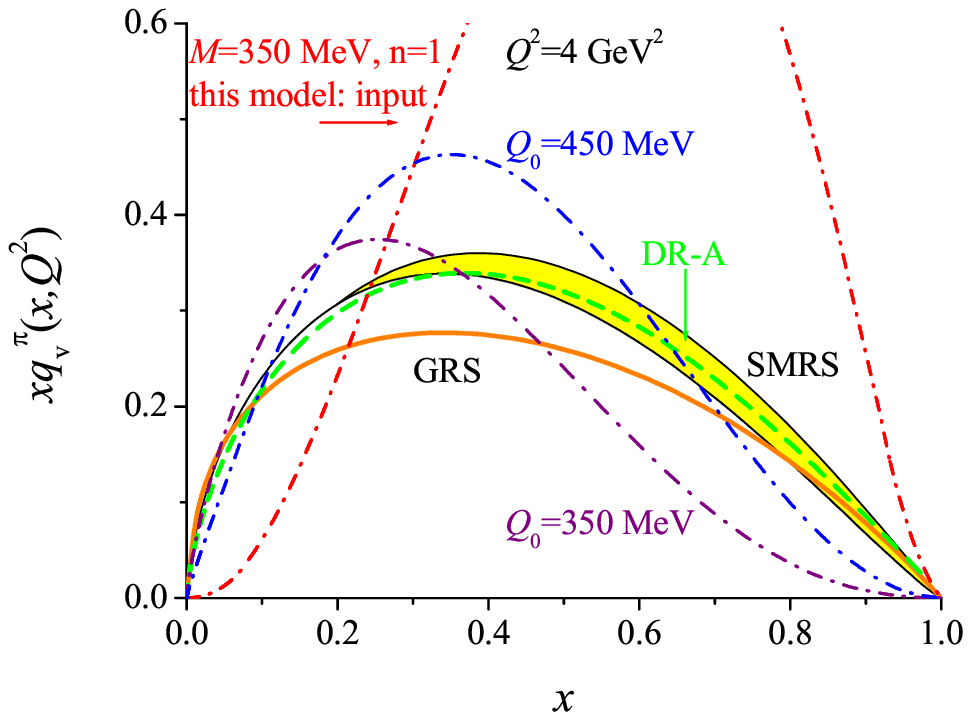} \\
\includegraphics[scale=0.8]{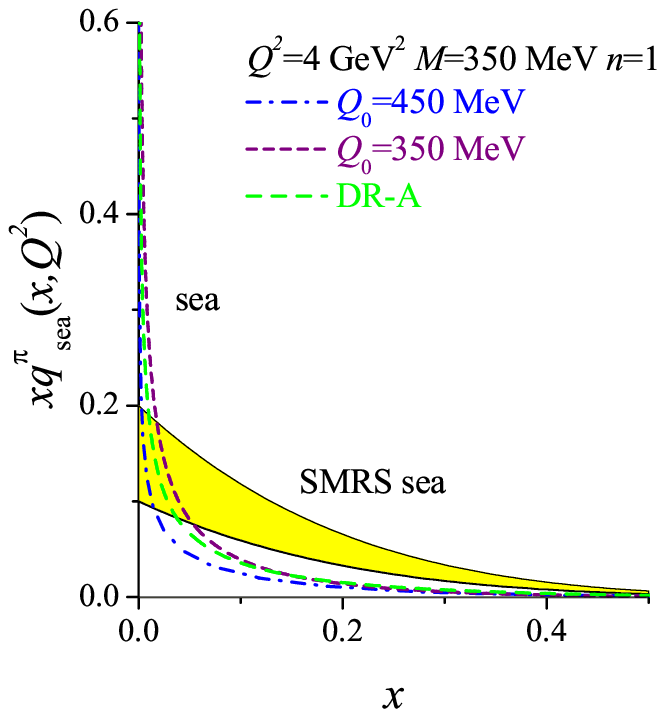} 
\includegraphics[scale=0.8]{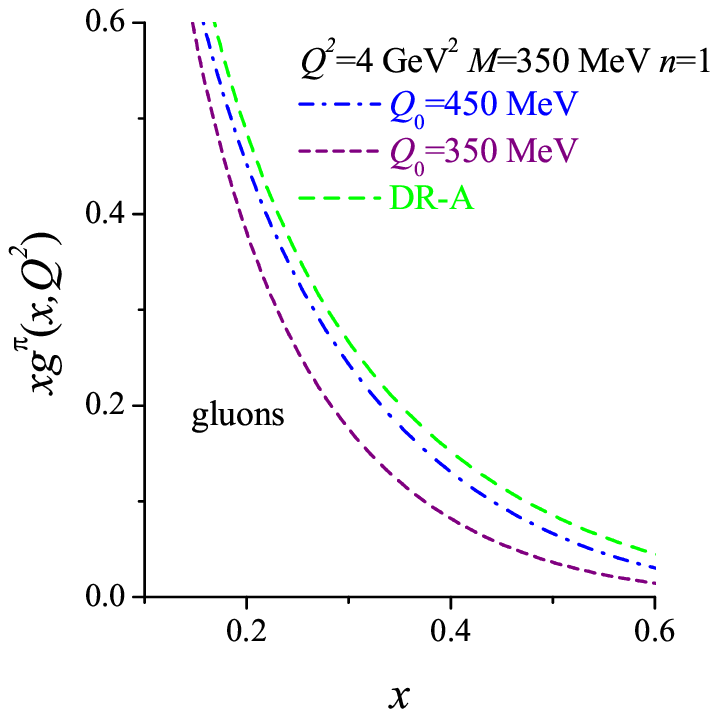} 
\end{center}
\caption{{\footnotesize $\pi^{+}$ valence, sea and gluon momentum
distributions. Upper panel: input distribution as calculated in
the model for $M=350$~MeV and $n=1$ (upper red dashed-dotted
curve) and after evolution to $Q^{2}=4$~GeV$^{2}$ from
$Q_{0}=450$~MeV (middle blue dashed-dotted curve) and 350 MeV
(lower purple dashed-dotted curve). Green dashed line represents
constant initial quark distribution \cite{ERA}. Experimental
parameterizations are depicted by a yellow band \cite{SMRS} and
orange line \cite{GRVS}. Lower panels: the same for
sea quarks and gluons with input distributions equal zero.}}%
\label{structureQ0}%
\end{figure}

Since, as will be discussed in Sect.\ref{sect:sum} , our model has
problems with the momentum sum rule, therefore we cannot use the
above prescription to fix $Q_{0}$. In Figs.\ref{structureQ0} we
simply show the shape of the valence, sea and gluon distributions
calculated in the model and evolved (in the leading log
approximation) to the scale $Q^{2}=4$~GeV$^{2}$ assuming
$Q_{0}=450$ and $350$~MeV. As initial conditions we take valence
quark distributions as calculated in the model with sea and gluon
distribution equal to zero at the initial scale $Q_{0}$. Therefore
both sea quarks and gluons are generated dynamically during the
evolution. In Figs.\ref{structureQ0} we also show ''experimental
data'' as extracted from the pion-proton Drell-Yan and direct
photon production \cite{SMRS,GRVS}. For comparison we also show
results of Ref.\cite{ERA} with a constant quark distribution at
initial scale.

It can be seen from Figs.\ref{structureQ0} that our quark
distribution differs from the distributions extracted from the
data. However, two existing experimental parameterizations are not
compatible. Interestingly, the constant initial quark distribution
after evolving to $Q^{2}=4$~GeV$^{2}$ fits very well
parameterization of Ref.\cite{SMRS}. Unfortunately for the sea,
both constant and our distributions do not follow the experimental
parameterization. This suggests that rather than compare model
results for pion parton distributions with the ones extracted from
the data it is perhaps more appropriate to calculate the cross
section itself and compare directly with the data.

There have been several other calculations of the pion structure
function in the literature. A result similar to ours was obtained
in Refs.\cite{Toki} where the authors calculate explicitly a
hand-bag diagram in a {\em local} NJL (nonbosonized) model in the
Bjorken limit. They, however, introduce an $x$ dependent cutoff in
order to regularize the divergent integrals and get proper
behavior of the valence quark distribution in the large $x$ limit.
Their result is in agreement with a similar calculation of
Ref.\cite{Heinzl}. In an approach based on Ward-Takahashi
identities, which is in fact equivalent to the local NJL model
with a sharp momentum cutoff, the valence quark distribution is
equal 1 over the whole range of $x$ \cite{ERA}. This very simple
quark distribution is properly normalized, in a sense that the
quark number is 1 and its total momentum is 1/2. The vanishing of
$q(x)$ for  $x \rightarrow 1$ is achieved by DGLAP evolution. We
have plotted the result of this evolution in
Fig.\ref{structureQ0}.

Direct calculations in the instanton model \cite{Dorokhov} show
phenomenologically quite similar behavior as ours. An advantage of
Refs.\cite{Dorokhov} is that they use well defined currents with
nonlocal pieces \cite{Birse}, whereas we use naive quark
billinears. This is reflected in wrong normalization of the first
moment of the quark distributions which at low scale should be 1
(for $\int dx \, x (u+\overline{d})$ ) as opposed to 0.93 what we get.
An even larger mismatch has been reported in a similar model of
Ref.\cite{HRS}, where the remainder of the momentum was attributed
to gluons and sea, which are, however, absent in our approach.

\begin{figure}[t]
\begin{center}
\includegraphics[scale=0.7]{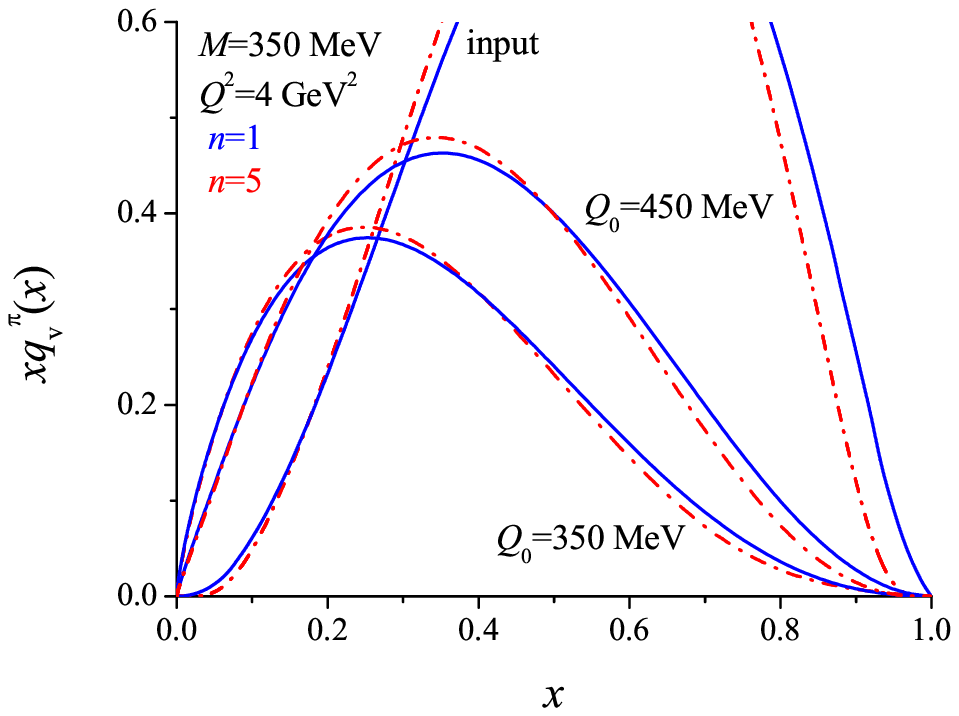}
\end{center}
\caption{{\footnotesize The $n$ dependence of valence quark
distributions for $M=350$~MeV and $n=1$ (solid) and 5
(dashed-dotted). Input density corresponds to the model result.
Two other sets of curves marked by $Q_0=450$ and $Q_0=350$~MeV
correspond to the distributions evolved to $Q^2=4$~GeV$^2$
assuming that the model scale was $Q_0$. }}
\label{ndep}%
\end{figure}

Our quark distributions show little $n$ dependence. This is depicted in Fig.
\ref{ndep}.

%
%
%
%

%

%
%
%
%

\section{Discussion}
\label{sect:sum}

In this paper we have calculated the Generalized Distribution
Amplitudes using the nonlocal, instanton motivated, chiral quark
model. The  nonlocality has been taken in the form of Eq.(\ref{M})
which allowed us to perform all calculations in the Minkowski
space. Introducing light cone integration variables allowed us to
perform most of the calculations analytically. As a result we were
left with the numerical integration in only one variable,
$dk_T^2$.

 We define the GDA's through the
matrix elements of the nonlocal quark billinears
(\ref{2piDA_def},\ref{spd_def}). Although in QCD this is certainly
a correct definition, one might envisage alternative definitions
which would be more appropriate for the nonlocal effective models
\cite{ERAPP,Dorokhov}. The reason is that in the limit when the
quark operators  are taken in the same point, quark billinears
(\ref{2piDA_def},\ref{spd_def})  do not correspond to the properly
normalized Noether currents. This is because in the nonlocal
models Noether currents get additional pieces which restore
Ward-Takahashi identities \cite{Birse}. It is not clear how to
generalize quark billinears (\ref{2piDA_def},\ref{spd_def}), since
such generalization is, in principle, process dependent. Therefore
an alternative way would be to calculate the whole scattering
amplitude directly in the effective model and then extract the
quantities one is interested in by imposing certain kinematical
constraints, like Bjorken limit for example. Although this
procedure seems at the first sight attractive, there is a problem
because the Bjorken limit requires large momentum transfer,
whereas the effective models are defined at low momenta.

In fact the precise determination of the normalization scale $Q_0$
at which the model is defined poses a serious problem. In the
instanton model of the QCD vacuum \cite{DP} that the pertinent
energy scale is of the order of the inverse instanton size
$Q_0=1/\rho\sim 600$~MeV. However, it has been argued in
Ref.\cite{ERA} that  $Q_0$ may be as low as 350~MeV. This estimate
is based on the requirement that the valence quark distribution
calculated in the model of Ref.\cite{ERA} and evolved from $Q_0$
to $Q=2$~GeV carry observed fraction of total momentum.
Unfortunately we cannot apply the same procedure in our case,
since the momentum sum rule is violated in our model. From both
equations, (\ref{2piDAI0_norm}) and (\ref{SPDI0_norm}) we find the
momentum fraction carried by quarks to be 93\%, independently of
$v$ and $\xi$. This causes the problem, as in the model we use,
the pion is built from constituent quarks (there are no gluons) so
there is 7\% of the pion momentum missing. A natural explanation
of this mismatch is that we missed some terms which, in the limit
where the quark operators are in the same point, would reduce
quark billinears (\ref{2piDA_def},\ref{spd_def}) to the full nonlocal
Noether currents \cite{Birse,BGR}.

\vspace{0.3cm}

\noindent\textbf{Acknowledgements}

We would like to thank W.~Broniowski and E.~Ruiz-Arriola for
comments and discussion. Special thanks are due to K.~Goeke and
all members of Inst. of Theor. Phys. II at Ruhr-University where
part of this work was completed. M.P. acknowledges discussion with
S.~Brodsky, A.~Dorokhov, V.Y.~Petrov, P.V.~Pobylitsa, M.~Polyakov,
A.~Radyushkin and Ch.~Weiss. A.R. acknowledges support of Polish
State Committee for Scientific Research under grant 2 P03B 048 22
and M.P. under grant 2 P03B 043 24.

\appendix
%
%

\section{Technical details of the calculation of the pion GDAs}

%
%
\label{appA}
%
%

Inserting (\ref{2pi_model}) and (\ref{skewed_model}) into (\ref{2piDA_def})
and (\ref{spd_def}) respectively we get
\begin{align}
\Phi_{2\pi}^{I=0}(u,v,s)  &  =\mathcal{J}_{1}(u,v,t)+\mathcal{J}%
_{2}(u,v,t)+\mathcal{J}_{3}(u,v,t),\label{2piI0} \nonumber \\
\Phi_{2\pi}^{I=1}(u,v,s)  &  =\mathcal{J}_{2}(u,v,t)-\mathcal{J}_{1}(u,v,t)
\end{align}
and
\begin{align}
H^{I=0}(X,\xi,t)  &  =\mathcal{I}_{1}(X,\xi,t)+\mathcal{I}_{2}(X,\xi
,t)+\mathcal{I}_{3}(X,\xi,t),\nonumber \\
H^{I=1}(X,\xi,t)  &  =\mathcal{I}_{1}(X,\xi,t)-\mathcal{I}_{2}(X,\xi,t).
\label{HI1}%
\end{align}
$\mathcal{J}_{1}$, $\mathcal{J}_{2}$, $\mathcal{J}_{3}$, $\mathcal{I}_{1}$,
$\mathcal{I}_{2}$, $\mathcal{I}_{3}$ stand for the integrals:
\begin{align}
\mathcal{J}_{1}(u,v,s)  &  =\frac{iN_{C}}{2(2\pi)^{4}F_{\pi}^{2}}\left.  \int
d^{2}k_{\bot}\,\int\limits_{-\infty}^{+\infty}dk_{-}\,\mathcal{T}%
_{2}(k-P,k-p_{1},k)\right|  _{\displaystyle k^{+}=uP^{+}},\label{J1}\\
\mathcal{J}_{2}(u,v,s)  &  =\frac{iN_{C}}{2(2\pi)^{4}F_{\pi}^{2}}\left.  \int
d^{2}k_{\bot}\,\int\limits_{-\infty}^{+\infty}dk_{-}\,\mathcal{T}%
_{2}(k-P,k-p_{2},k)\right|  _{\displaystyle k^{+}=uP^{+}},\label{J2}\\
\mathcal{J}_{3}(u,v,s)  &  =\frac{iN_{C}}{2(2\pi)^{4}F_{\pi}^{2}}\left.  \int
d^{2}k_{\bot}\,\int\limits_{-\infty}^{+\infty}dk_{-}\,\mathcal{T}%
_{1}(k-P,k)\right|  _{\displaystyle k^{+}=uP^{+}}, \label{J3}%
\end{align}%
\begin{align}
\mathcal{I}_{1}(X,\xi,t)  &  =\frac{iN_{C}}{4\left(  2\pi\right)  ^{4}F_{\pi
}^{2}}\left.  \int d^{2}k_{\bot}\,\int\limits_{-\infty}^{+\infty}%
dk_{-}\,\mathcal{T}_{2}\left(  k-\frac{\Delta}{2},k-\bar{p},k+\frac{\Delta}%
{2}\right)  \right|  _{\displaystyle k^{+}=X\bar{p}^{+}},\label{I1}\\
\mathcal{I}_{2}(X,\xi,t)  &  =\frac{iN_{C}}{4\left(  2\pi\right)  ^{4}F_{\pi
}^{2}}\left.  \int d^{2}k_{\bot}\,\int\limits_{-\infty}^{+\infty}%
dk_{-}\,\mathcal{T}_{2}\left(  k-\frac{\Delta}{2},k+\bar{p},k+\frac{\Delta}%
{2}\right)  \right|  _{\displaystyle k^{+}=X\bar{p}^{+}},\label{I2}\\
\mathcal{I}_{3}(X,\xi,t)  &  =\frac{iN_{C}}{4\left(  2\pi\right)  ^{4}F_{\pi
}^{2}}\left.  \int d^{2}k_{\bot}\,\int\limits_{-\infty}^{+\infty}%
dk_{-}\,\mathcal{T}_{1}\left(  k-\frac{\Delta}{2},k+\frac{\Delta}{2}\right)
\right|  _{\displaystyle k^{+}=X\bar{p}^{+}}, \label{I3}%
\end{align}
with $\mathcal{T}_{1}$ and $\mathcal{T}_{2}$ defined in (\ref{T1}, \ref{T2}). 
It is convenient to introduce the scaled variables:
\begin{equation}
\kappa^{\mu}=\frac{k^{\mu}}{\Lambda},\qquad\omega^{2}=\frac{s}{\Lambda^{2}%
},\qquad\vec{\tau}_{\perp}=\frac{\vec{p}_{\perp}}{\Lambda},\qquad\tau
=\frac{\tilde{t}}{4\Lambda^{2}},\qquad\vec{\delta}_{\perp}=\frac{\vec{\Delta
}_{\perp}}{2\Lambda}%
\end{equation}
and the notation:
\begin{equation}
\bar{u}=1-u,\qquad\bar{v}=1-v.
\end{equation}
We introduce the factors
\begin{equation}
f_{i}=\prod\limits_{{}_{\scriptstyle k\neq i}^{\scriptstyle k=1}}^{4n+1}%
\frac{1}{z_{i}-z_{k}}, \label{fidef}%
\end{equation}
which we will use below. $z_{i}$ are $4n+1$ roots of the Eq.
(\ref{G}). The factors $f_{i}$ have a property:
\begin{equation}
\sum\limits_{i=1}^{4n+1}z_{i}^{m}f_{i}=\left\{
\begin{array}
[c]{ccc}%
0 & \mbox{for} & m<4n,\\
1 & \mbox{for} & m=4n.
\end{array}
\right.  \label{fiprop}%
\end{equation}
This property is crucial for the convergence of the integrals (\ref{J1}%
-\ref{I3}), in analogy to the case of the pion distribution amplitude
\cite{PR}. This property is true for any set of different $4n+1$ numbers,
irrespectively to the fact that they are solutions of certain polynomial
equation. \newline It can be shown that
\begin{equation}
\mathcal{J}_{3}(u,v,s)=\mathcal{J}_{3}(u,s)=-\mathcal{J}_{3}(\bar{u},s),
\label{j3_u_sym}%
\end{equation}%
\begin{equation}
\mathcal{J}_{1}(\bar{u},v,s)=-\mathcal{J}_{2}(u,v,s), \label{j1_u_sym}%
\end{equation}%
\begin{equation}
\mathcal{J}_{1}(u,\bar{v},s)=\mathcal{J}_{2}(u,v,s) \label{j1_v_sym}%
\end{equation}
and
\begin{equation}
\mathcal{I}_{2}(-X,\xi,t)=-\mathcal{I}_{1}(X,\xi,t),\qquad\mathcal{I}%
_{3}(-X,\xi,t)=-\mathcal{I}_{3}(X,\xi,t), \label{X_sym}%
\end{equation}%
\begin{equation}
\mathcal{I}_{2}(X,-\xi,t)=\mathcal{I}_{1}(X,\xi,t),\qquad\mathcal{I}%
_{3}(X,-\xi,t)=\mathcal{I}_{3}(X,\xi,t). \label{xi_sym}%
\end{equation}
The properties (\ref{j3_u_sym}-\ref{xi_sym}), together with Eqs.
(\ref{2piI0}-\ref{HI1}), provide the correct symmetry properties
for $2\pi$DAs
(\ref{I02piDA_sym}, \ref{I12piDA_sym}) and SPDs (\ref{HI0_sym}, \ref{HI1_sym}%
)\newline For $\mathcal{J}_{3}(u,s)$ we get analytical results:
\begin{equation}
\mathcal{J}_{3}(u,s)=\frac{N_{C}M^{2}}{\left(  2\pi\right)  ^{2}F_{\pi
}^{2}}\sum_{i=1}^{4n+1}\sum_{k=1}^{4n+1}z_{i}^{n}z_{k}^{n}\,f_{i}%
\,f_{k}\left(  uz_{i}^{2n}-\bar{u}z_{k}^{2n}\right)  \ln\left(  -u\bar
{u}\omega^{2}+1+\bar{u}z_{i}+uz_{k}\right)  . \label{J3_res}%
\end{equation}
If $0\leq u\leq\bar{v}$ then $\mathcal{J}_{2}(u,v,s)$ reads
\begin{align}
\lefteqn{\mathcal{J}_{2}(u,v,s)}\nonumber\\
&  =(-1)^{n+1}\frac{iN_{C}M^{2}}{\left(  2\pi\right)  ^{3}F_{\pi}^{2}}%
\sum_{i=1}^{4n+1}z_{i}^{n}f_{i}\int\limits_{0}^{\infty}d\left(  \kappa_{\bot
}^{2}\right)  \,\frac{u^{7n+1}\left[  \left(  -u\bar{u}\omega^{2}+\vec{\kappa
}_{\perp}^{2}+1+\bar{u}z_{i}\right)  \right]  ^{n}}{\prod\limits_{k=1}%
^{4n+1}\left(  -u\bar{u}\omega^{2}+\vec{\kappa}_{\perp}^{2}+1+\bar{u}%
z_{i}+uz_{k}\right)  }\nonumber\\
&  \times\int\limits_{C(0,1)}d\lambda\,\frac{\lambda^{4n}}{\prod
\limits_{k=1}^{4n+1}\left(  A(u)\lambda^{2}+B_{ik}(u,v)\lambda+A(u)\right)
}\,g(u,v)\nonumber\\
=:  &  \mathcal{F}(u,v,s), \label{J2_u<vb}%
\end{align}
where
\begin{align}
\lefteqn{g(u,v)=\mu^{2}\left[  -\bar{u}\left(  \frac{-u\bar{u}\omega^{2}%
+\vec{\kappa}_{\perp}^{2}+1+\bar{u}z_{i}}{u}\right)  ^{2n}+uz_{i}^{2n}\right.
}\nonumber\\
&  \left.  -\left(  u-\bar{v}\right)  \left(  \frac{A(u)\lambda^{2}%
+b_{i}(u,v)\lambda+A(u)}{\lambda u}\right)  ^{2n}\right] \nonumber\\
&  +z_{i}^{2n}\left(  \frac{-u\bar{u}\omega^{2}+\vec{\kappa}_{\perp}%
^{2}+1+\bar{u}z_{i}}{u}\right)  ^{2n}\left(  \frac{A(u)\lambda^{2}%
+b_{i}(u,v)\lambda+A(u)}{\lambda u}\right)  ^{2n}\nonumber\\
&  \times\left[  -vu\bar{u}\omega^{2}+\underbrace{\left(  u+\bar{u}-v\right)
}_{\bar{v}}\kappa_{\bot}^{2}+\bar{u}\left(  1+z_{i}\right)  +\left(  u-\bar
{u}\right)  \sqrt{\kappa_{\bot}^{2}\tau_{\bot}^{2}}\frac{\lambda^{2}%
+1}{2\lambda}\right]  . \label{g}%
\end{align}
If $\bar{v}\leq u\leq1$, then
\begin{equation}
\mathcal{J}_{2}(u,v,s)=-\mathcal{F}(\bar{u},\bar{v},s),
\end{equation}
with $\mathcal{F}(u,v,s)$ defined in (\ref{J2_u<vb}). The symbols
$A(u)$, $b_{i}(u,v)$, $B_{ik}(u,v)$ in Eq. (\ref{J2_u<vb},
\ref{g}) stand for
\[
 A(u)=u\sqrt{\kappa_{\bot}^{2}\tau_{\bot}^{2}},
\]%
\[
b_{i}(u,v)=u\tau_{\perp}^{2}+uv(u-\bar{v})\omega^{2}+\bar{v}(\vec{\kappa
}_{\perp}^{2}+1)-(u-\bar{v})z_{i},
\]%
\[
B_{ik}(u,v)=u\tau_{\perp}^{2}+uv(u-\bar{v})\omega^{2}+\bar{v}(\vec{\kappa
}_{\perp}^{2}+1)-(u-\bar{v})z_{i}+uz_{k}.
\]

\noindent Because $H^{I=0}$ and $H^{I=1}$ are symmetric in $\xi$ we can assume
$\xi\geq0$. Then $\mathcal{I}_{1}(X, \xi, t)$ is nonzero only if $-\xi\leq X
\leq1$. Similarly $\mathcal{I}_{3}(X, \xi, t)$ is nonzero only if $-\xi\leq X
\leq\xi$. \newline For $-\xi\leq X \leq\xi$, $\mathcal{I}_{3}(X, \xi, t)$ and
$\mathcal{I}_{1}(X, \xi, t)$ read
\begin{align}
&  \mathcal{I}_{3} (X, \xi,t)\nonumber\\
&  = (-1)^{n} \frac{i N_{C} M^{2}} {2 (2 \pi)^{3} F_{\pi}^{2}}
\int\limits_{0}^{\infty} d \left(  \kappa_{\bot}^{2}\right)  \, \sum
_{i=1}^{4n+1} f_{i} \int\limits_{C(0,1)} d\lambda\, \left[  (X + \xi)
\lambda\, z_{i} \, \left(  A \lambda^{2} + b_{i} \lambda+ A\right)  \right]
^{n}\nonumber\\
&  \times\frac{(X + \xi) \left[  \left(  X+\xi\right)
\lambda\,z_{i}\right] ^{2n} + (X - \xi) \left(  A \lambda^{2} +
b_{i} \lambda+ A \right)  ^{2n}} {\prod\limits_{k=1}^{4n+1} \left[
A \lambda^{2} + B_{ik} \lambda+ A\right] },\label{I3_a} %
\end{align}%
\begin{align}
\lefteqn{ \mathcal{I}_{1} (X, \xi,t) = (-1)^{n} \frac{i N_{C} M^{2}} {2 (2
\pi)^{3} F_{\pi}^{2}} \int\limits_{0}^{\infty} d\left(  \kappa_{\bot}%
^{2}\right)  \, \sum_{i=1}^{4n+1} f_{i} \, }\nonumber\\
&  \times\int\limits_{C(0,1)} d \lambda\frac{ [(X + \xi)
\lambda]^{7n+1} z_{i}^{n} \left(  A \lambda^{2} + b_{i} \lambda+
A\right)  ^{n}} {\prod\limits_{k=1}^{4n+1} \left[  \left(  A
\lambda^{2} + B_{ik} \lambda+ A \right)  \left(  C \lambda^{2} +
D_{ik} \lambda+ C \right)  \right]  } \, h_{1}^{(a)} (X,
\xi,t),\nonumber
\end{align}
where
\begin{align}
&  h_{1}^{(a)} (X, \xi,t)\nonumber\\
&  = \mu^{2} \left[  (X - \xi) \left(  \frac{A \lambda^{2} + b_{i} \lambda+ A}
{(X + \xi) \lambda} \right)  ^{2n} + (X + \xi) z_{i}^{2n} - (X - 1) \left(
\frac{C \lambda^{2} + d_{i} \lambda+ C} {(X + \xi) \lambda} \right)  ^{2n}
\right] \nonumber\\
&  + z_{i}^{2n} \left(  \frac{A \lambda^{2} + b_{i} \lambda+ A} {(X + \xi)
\lambda} \right)  ^{2n} \left(  \frac{C \lambda^{2} + d_{i} \lambda+ C} {(X +
\xi) \lambda} \right)  ^{2n}\nonumber\\
&  \times\left[  \left(  \xi^{2}-X^{2}\right)  (1-\xi)\tau+\left(
\xi+1\right)  \kappa_{\bot}^{2}+X\sqrt{\kappa_{\bot}^{2}\delta_{\bot}^{2}%
}\frac{\lambda^{2}+1}{\lambda}+\left(  \xi-1\right)  \delta_{\bot}^{2} + (\xi-
X) (1 + z_{i}) \right]  .\nonumber\\
\end{align}
For $\xi\leq X \leq1$, $\mathcal{I}_{1} (X, \xi,t)$ reads
\begin{align}
\lefteqn{\mathcal{I}_{1} (X, \xi,t) = (-1)^{n+1} \frac{i N_{C} M^{2}} {2
(2 \pi)^{3} F_{\pi}^{2}} \int\limits_{0}^{\infty} d\left(  \kappa_{\bot}%
^{2}\right)  \, \sum_{i=1}^{4n+1} f_{i} }\nonumber\\
&  \times\int\limits_{C(0,1)} d\lambda\frac{[(X - 1) \lambda]^{6n+1} \left(  C
\lambda^{2} + f_{i} \lambda+ C \right)  ^{n} \left(  C \lambda^{2} + g_{i}
\lambda+ C \right)  ^{n}} { \prod\limits_{k=1}^{4n+1} \left[  \left(  C
\lambda^{2} + F_{ik} \lambda+ C \right)  \left(  C \lambda^{2} + G_{ik}
\lambda+ C \right)  \right]  } \, h_{1}^{(b)} (X, \xi, t),\nonumber\\
\end{align}
where
\begin{align}
&  h_{1}^{(b)} (X, \xi, t)\nonumber\\
&  = \mu^{2} \left[  (X - \xi) \left(  \frac{C \lambda^{2} + g_{i} \lambda+ C}
{(X - 1) \lambda} \right)  ^{2n} + (X + \xi) \left(  \frac{C \lambda^{2} +
f_{i} \lambda+ C} {(X-1) \lambda} \right)  ^{2n} \right. \nonumber\\
&  \left.  - (X - 1) z_{i}^{2n} \right] \nonumber\\
&  + \left(  \frac{C \lambda^{2} + f_{i} \lambda+ C} {(X - 1) \lambda}
\right)  ^{2n} \left(  \frac{C \lambda^{2} + g_{i} \lambda+ C} {(X - 1)
\lambda} \right)  ^{2n} z_{i}^{2n}\nonumber\\
&  \times\left[  \frac{\xi^{2}-1}{X-1}\kappa_{\bot}^{2}+\xi\sqrt{\kappa_{\bot
}^{2}\delta_{\bot}^{2}}\frac{\lambda^{2}+1}{\lambda}+\left(  X-1\right)
\delta_{\bot}^{2}+\frac{\xi^{2}-X^{2}}{X-1}(1+z_{i}) \right]  .\label{hb1}
\end{align}
The symbols $A$, $b_{i}$, $B_{ik}$, $C$, $d_{i}$, $D_{ik}$, $f_{i}$, $F_{ik}$,
$g_{i} $, $G_{ik}$ in Eqs. (\ref{I3_a} - \ref{hb1}) stand for
\[
A = 2 X \sqrt{ \kappa_{\bot}^{2} \delta_{\bot}^{2} },
\]%
\[
b_{i} = 2 \xi\left[  \left(  X^{2} - \xi^{2} \right)  \tau+ \kappa_{\bot}^{2}
+ \delta_{\bot}^{2} + 1 \right]  + (\xi- X) z_{i},
\]%
\[
B_{ik} = 2 \xi\left[  \left(  X^{2} - \xi^{2} \right)  \tau+ \kappa_{\bot}^{2}
+ \delta_{\bot}^{2} + 1 \right]  + (\xi- X) z_{i} + (\xi+ X) z_{k},
\]%
\[
C = (1 - X) \sqrt{ \kappa_{\bot}^{2} \delta_{\bot}^{2} },
\]%
\[
d_{i} = (X - 1) \left[  (\xi+ X) (1 - \xi) \tau+ \delta_{\bot}^{2} \right]  -
(1 + \xi) \left(  \kappa_{\bot}^{2} + 1 \right)  + (X - 1) z_{i},
\]%
\[
D_{ik} = (X - 1) \left[  (\xi+ X) (1 - \xi) \tau+ \delta_{\bot}^{2} \right]  -
(1 + \xi) \left(  \kappa_{\bot}^{2} + 1 \right)  + (X - 1) z_{i} - (\xi+ X)
z_{k},
\]%
\[
f_{i} = (X - 1) \left[  (\xi+ X) (1 - \xi) \tau+ \delta_{\bot}^{2} \right]  -
(1 + \xi) \left(  \kappa_{\bot}^{2} + 1 \right)  - (\xi+ X) z_{i},
\]%
\[
F_{ik} = (X - 1) \left[  (\xi+ X) (1 - \xi) \tau+ \delta_{\bot}^{2} \right]  -
(1 + \xi) \left(  \kappa_{\bot}^{2} + 1 \right)  - (\xi+ X) z_{i} + (X - 1)
z_{k},
\]%
\[
g_{i} = (X - 1) \left[  (\xi- X) (1 + \xi) \tau- \delta_{\bot}^{2} \right]  +
(1 - \xi) \left(  \kappa_{\bot}^{2} + 1 \right)  - (\xi- X) z_{i},
\]%
\[
G_{ik} = (X - 1) \left[  (\xi- X) (1 + \xi) \tau- \delta_{\bot}^{2} \right]  +
(1 - \xi) \left(  \kappa_{\bot}^{2} + 1 \right)  - (\xi- X) z_{i} - (X - 1)
z_{k}.
\]

%
%
\end{document}